\documentclass[12pt,titlepage]{article}
\usepackage[utf8]{inputenc}
\usepackage[margin=1in,letterpaper]{geometry}
\usepackage[nodisplayskipstretch]{setspace}\doublespace

\usepackage{mathtools,amssymb,microtype,bm,dsfont,graphicx,color,soul,placeins,booktabs,multicol,multirow,rotating,natbib}
\usepackage{adjustbox,courier,soul,listings}

\usepackage{enumitem}
\newlist{tabitem}{itemize}{1}
\setlist[tabitem]{wide=0pt,nosep,leftmargin=*,label=\textbullet,after=\vspace{-\baselineskip},before=\vspace{-0.6\baselineskip}\csname par\endcsname\raggedright}

\usepackage{titling}\thanksmarkseries{fnsymbol}
\pretitle{\begin{center}\Large}
\preauthor{\begin{center}\normalsize\lineskip 1em \begin{tabular}[t]{c}}
\predate{\begin{center}\normalsize}

\usepackage{titlesec}
\titleformat{\subsubsection}[runin]{\normalfont\itshape}{\thesubsubsection}{1em}{}[.\enspace]
\titlespacing{\subsubsection}{\parindent}{\parskip}{0pt}

\definecolor{green}{rgb}{0.0,0.47,0.44}
\usepackage{marginnote}
\newcounter{ctrpend}

\bibpunct[, ]{(}{)}{,}{a}{}{,}

\setcitestyle{notesep={:}}\newcommand{\p}[1]{#1}\newcommand{\pp}[1]{#1}

\usepackage[breaklinks,hidelinks]{hyperref}

\usepackage{caption}
\newcommand{\codefont}{\fontsize{10pt}{12pt}\selectfont\ttfamily}
\lstnewenvironment{code}[2]{\singlespace\lstset{basicstyle=\codefont#2, breaklines=true, breakatwhitespace=true, breakindent=0pt, frame=none} \noindent\codefont \textbf{\ul{#1}}}{}

\makeatletter
\def\@seccntformat#1{\@ifundefined{#1@cntformat}{\csname the#1\endcsname\quad}{\csname #1@cntformat\endcsname}}
\let\oldappendix\appendix
\renewcommand\appendix{\oldappendix\newcommand{\section@cntformat}{\appendixname~\thesection:\ }}
\makeatother

\newcommand{\TITLE}{Artificial Intelligence and Strategic Decision-Making:\\Evidence from Entrepreneurs and Investors}

\title{\TITLE\thanks{Authors are listed in alphabetical order and contributed equally.  We thank Ruijing Chen, Can He, and Hazel Lim for excellent research assistance, and thank our editor and referees, as well as participants at the Academy of Management Meeting 2023, Theory-Based View in Strategy \& Entrepreneurship Conference, AI \& Strategy Consortium 2024, AI \& Strategy Workshop 2024, Ion Management Science Lab at Bocconi University, University of Chile, University of Los Andes, and DRUID 2024 for valuable comments.  We are grateful to our collaborating organizations and to the University of Michigan, Bocconi University, and the INSEAD eLab Research Fund for financial support.  This experiment was approved by the INSEAD IRB office and pre-registered at the AEA RCT Registry (AEARCTR-0011942).}}
\author{Felipe A. Csaszar \\ University of Michigan \\ \texttt{fcsaszar@umich.edu} \\[2.5ex] Harsh Ketkar \\ University of Texas at Austin \\ \texttt{harsh.ketkar@mccombs.utexas.edu} \\[2.5ex] Hyunjin Kim \\ INSEAD \\ \texttt{hyunjin.kim@insead.edu}}
\date{{\vskip 2em}August 1, 2024}

\begin{document}
\begin{onehalfspace}\maketitle
\begin{abstract}
This paper explores how artificial intelligence (AI) may impact the strategic decision-making (SDM) process in firms.  We illustrate how AI could augment existing SDM tools and provide empirical evidence from a leading accelerator program and a startup competition that current Large Language Models (LLMs) can generate and evaluate strategies at a level comparable to entrepreneurs and investors.  We then examine implications for key cognitive processes underlying SDM---search, representation, and aggregation.  Our analysis suggests AI has the potential to enhance the speed, quality, and scale of strategic analysis, while also enabling new approaches like virtual strategy simulations.  However, the ultimate impact on firm performance will depend on competitive dynamics as AI capabilities progress.  We propose a framework connecting AI use in SDM to firm outcomes and discuss how AI may reshape sources of competitive advantage.  We conclude by considering how AI could both support and challenge core tenets of the theory-based view of strategy.  Overall, our work maps out an emerging research frontier at the intersection of AI and strategy.

\end{abstract}
\end{onehalfspace}

\section{Introduction}

\subsection{Could AI make strategic decisions?}

Strategic decision-making (SDM) involves identifying and choosing among long-term courses of action that can determine firm performance.  SDM is inherently challenging due to its high-stakes, unstructured and ambiguous nature, novel and complex characteristics with multiple parts and interactions, delayed and noisy feedback, and potential to overwhelm the bounded cognitive resources of managers (\citealt{Csas24}).  Traditionally, making strategic decisions has been a purely human endeavor, relying on the expertise, intuition, and cognitive abilities of experienced strategists.  However, rapid advancements in artificial intelligence (AI) may be beginning to challenge this paradigm.  AI has progressively made strides in aspects of decision-making that were once thought to be exclusively human domains.  From mastering complex strategy games like Go, StarCraft, and Diplomacy \citep{Risi20} to excelling at other decision-making tasks such as prediction \citep{Agra18,Cowg19,Kim24}, creative writing \citep{Noy23,Dosh23}, business ideas and advice \citep{Bryn23,Giro23,Dell23,Otis23}, and scientific problem-solving \citep{Boik23,Ludw24,Mann24}, AI has continually managed to match or surpass human experts in various fields.  Given these advancements, it is reasonable to consider that AI may increasingly play a significant role in the generation and evaluation of strategies.

To understand how AI may shape SDM, it is useful to examine the evolution of financial decision-making over the past three decades.  Initially, human traders made all trading decisions, but today, algorithms account for over 78\% of trading decisions \citep{SEC20}.  Algorithmic trading has surpassed human trading due to its reliance on quantitative inputs that can be efficiently processed and analyzed by predictive algorithms, resulting in better, faster, and more cost-effective trades.  However, unlike financial trading, which often involves rule-based decisions, SDM requires nuanced judgment and interpretation, and heavily depends on open-ended, qualitative textual inputs and outputs.  Strategists consider various inputs such as news, user stories, and market reports, and produce outputs like strategic plans, memos, and speeches.  Until recently, computers have struggled to handle this type of unstructured textual data, making SDM largely inaccessible to AI---or, at best, achievable
only in the distant future.

The advent of Large Language Models (LLMs) has made AI-augmented SDM seem more plausible, potentially accelerating this timeline.  LLMs are word prediction algorithms trained on a large corpus of text documents.  Surprisingly, when sufficiently complex LLMs are trained on sufficiently large corpora, LLMs start exhibiting ``emergent'' capabilities---such as the ability to answer questions, summarize, and reason logically---that are not present with simpler models or smaller corpora \citep{Wei22}.  The capabilities of LLMs make AI-augmented SDM more plausible for three reasons.  First, LLMs can deal with the type of textual data comprising the typical inputs and outputs of strategy.  Second, LLMs have matched or surpassed human performance in tasks that demand reasoning skills akin to those of strategists, such as successfully passing professional-level exams in fields like medicine, psychology, and law \citep{Bube23,Katz23}.  Finally, the corpora used to train LLMs include valuable information for SDM, such as consumer preferences, competitor information, and strategy knowledge \citep{Hort23,Bran23}.  Consequently, the possibility that AI could be used to make strategic decisions---those high-stakes, complex, and uncertain decisions that managers, consultants, and MBA students face while determining the firm's strategy---seems increasingly plausible.

However, significant uncertainty remains regarding how AI-augmented SDM may look and how well it may perform.  This uncertainty stems from several concerns, especially those that arise from key aspects of the theory-based view of strategy.  First, LLMs are fundamentally designed for predicting next words and generating fluent text, which may limit their capacity for creating forward-looking causal theories and strategies \citep{Zell23}.  Second, since LLMs rely on past data, there is a risk of reproducing conventional and widely accepted strategies, potentially reducing the novelty, diversity, and uniqueness of strategic ideas \citep{Feli24}.  Third, the generality of LLMs may limit their effectiveness in specific strategic decisions, as LLMs might not adequately capture the unique needs and contexts of individual firms.  Moreover, since SDM has been exclusively a human endeavor, the emergence of AI-augmented SDM creates uncertainty regarding how SDM processes and outcomes may evolve.

Given these opportunities and uncertainties, our paper seeks to address several critical questions arising from this new paradigm: (1) How may AI-augmented SDM look?  (2) How effective are current LLMs at making strategic decisions?  (3) What are the implications of using AI in SDM?  By examining these questions, we aim to provide a clearer picture of the potential and challenges associated with integrating AI into SDM.

\subsection{Our approach and contribution}

To answer these questions, we proceed as follows.  First, to illustrate how AI-augmented SDM may look, we examine current SDM tools such as Porter's Five Forces, Devil's Advocate, and others, and reimagine ``AI-augmented'' versions of them.  We present these reimagined tools in vignettes to demonstrate their potential applications.  These vignettes provide diverse illustrations of what AI-augmented strategy might look like, employing existing SDM tools used in business education and practice.

Second, to measure the effectiveness of AI-augmented SDM, we provide quantitative evidence through two studies that compare humans and LLMs in their ability to \emph{generate} and \emph{evaluate} forward-looking entrepreneurial strategies.  The first study, in collaboration with a startup accelerator, experimentally compares LLM-generated entrepreneurial business plans to those by entrepreneurs seeking venture capital.  The second study uses data from a startup competition to compare LLM evaluations of business plans with those of venture capital and angel investors.  These studies provide empirical evidence on how LLMs can enhance SDM in realistic strategy environments with high uncertainty, hence complementing the emerging literature on business uses of LLMs, which has looked at either tasks that do not involve SDM (e.g., writing and creativity; \citealt{Noy23,Giro23,Dosh23}) or more stylized versions of it (e.g., internal consulting tasks and small-business mentoring; \citealt{Dell23,Otis23}).

Third, to examine the implications of using AI in SDM, we look at the cognitive processes that underlie SDM \citep{Csas22}---search, representation, and aggregation---and examine how AI may affect them.  This allows us to theorize and develop scenarios regarding how AI may influence the quality and heterogeneity of firms' strategies \citep{Levi97}, the complexity of strategic representations \citep{Csas20b}, the role of business experiments \citep{Camu20,Koni22}, and the importance of theories in strategy \citep{Feli17}.

Our work provides several findings.  First, the vignettes demonstrate the feasibility of using current LLMs to automate commonly used SDM tools.  Second, our empirical studies show that LLMs exhibit the ability to generate and evaluate strategic ideas at a level comparable to that of entrepreneurs and investors in realistic contexts.  Third, our analyses indicate that many important questions in strategy are closely tied to the advancement of AI, with the evolution of strategy theory and practice significantly depending on its progress.  For example, we show how the nature of competitive advantage may change based on AI's future capabilities: it could remain Ricardian (based on unique resources), become Schumpeterian (driven by innovation), or potentially cease to exist altogether.  Additionally, we discuss how AI's impact on strategy may be constrained by some of the tenets of the theory-based view of strategy \citep{Feli09}, or conversely, how AI could potentially expand this theoretical framework.  These considerations underscore the profound influence AI developments may have on the strategy field.

A more general contribution of our work is to introduce a framework to understand how AI might change SDM.  This framework examines the fundamental antecedents and consequences of AI use in strategic decision-making (SDM) by analyzing the cognitive processes it changes---search, representation, and aggregation of strategies---and the outcomes it affects---generation and evaluation of strategies---and linking these SDM outcomes to competitive outcomes like value creation and profits.  This way of understanding how AI affects strategy complements prior work that has conceptualized AI as decreasing firms' cost of making better predictions \citep{Agra18,Cowg19,Alle22,Kim24} and increasing worker productivity across a variety of tasks \citep{Noy23,Bryn23,Giro23,Dosh23,Dell23,Jia24}.  Given AI's potential to expand the limits of bounded rationality---a crucial factor in SDM---there are numerous potential applications and research opportunities.  Our framework serves as a map to navigate this emerging landscape.

The next section explores how AI-augmented SDM may look.  Section~\ref{sec:evidence} provides empirical evidence on how AI can generate and evaluate strategies, leveraging data from a leading accelerator program and a startup competition.  Section~\ref{sec:implications} examines how AI-augmented SDM may evolve by looking at the changes in underlying processes (search, representation, and aggregation).  Section~\ref{sec:conclusion} concludes with broader implications, including impacts on competition, the theory-based view, and opportunities for strategy research.

\section{How may AI-augmented SDM look?}\label{sec:how}

To understand how AI may affect SDM, we first explore how AI could be incorporated into SDM tools that are already used to make strategic decisions.  These tools are used by managers to generate and/or evaluate strategic ideas.\footnote{Understanding problem-solving in terms of generation and evaluation is rooted in Newell and Simon's \citeyearpar[\p 121]{Newe76} classic model of problem-solving, which involves generating potential solutions and evaluating them to determine their adequacy.  It is important to note that generation processes have a scope, which determines the size of the problem space.  For example, when generating a strategy for a firm, one could ask for a strategy at the corporate, division, or product level.  Even at the product level, the scope may be reduced if certain attributes of the product cannot be changed.}  For instance, managers use scenario planning \citep{Scho95} to generate possible strategies, and frameworks like VRIN \citep{Barn91} or the Five Forces \citep{Port79b} to evaluate strategies.

In this section we re-imagine how four popular SDM tools---Scenario Planning, Porter's Five Forces, Devil's Advocate, and Wisdom of the Crowd---could be augmented by using AI.  Although SDM tools are inherently incomplete and inaccurate, as all representations are \citep[\p 17]{Simo47}, and cannot encompass the full range of strategic problems managers face, it is valuable to discuss how these tools might be enhanced by AI.  This discussion serves two primary purposes in our paper: first, to offer concrete examples that anchor our analyses, and second, to establish a foundation for later discussions on more advanced AI applications in SDM.  Most of the examples below use business education as a setting, as this context is likely to be familiar to our readers.  The LLM used to generate all the answers in this section is GPT-4 \citep{Open23}.

\subsection{Scenario planning}

We asked the LLM to conduct scenario planning, a tool used for strategy generation, by making the following request (see Appendix~\ref{app:scenario-planning} for details, including full prompts): ``In one sentence, describe three events that could make business schools obsolete, or dramatically decrease their demand.''  The LLM generated three possible scenarios: (1) advanced AI allows more customizable and interactive business education than traditional business schools, (2) employer preferences shift towards practical skills, reducing the value of business school education, and (3) students prefer micro-credentials and online courses over business school degrees.  All are reasonable scenarios that address some of the most frequently acknowledged concerns in the industry.

As a second stage of the scenario planning, we asked the LLM to devise a plan for a business school to thrive in scenario~1, the emergence of advanced AI that allows for personalized education.  In response, the LLM provided a six-point plan of what a business school could do in this scenario (Appendix~\ref{app:scenario-planning}).  Some of the recommendations included setting up partnerships with AI companies, as well as upskilling faculty to incorporate AI in their teaching.  Interestingly, even without any specific references in the prompt, the LLM was able to identify a key secondary stakeholder in the industry---AI companies---that a human decision-maker could have missed.

It is worth noting that most of the generated scenarios as well as the subsequent recommendations are qualitatively similar to what an MBA student could produce, and that one can request the LLM to produce an arbitrary number of scenarios.  Although at some point some repetition starts to appear, GPT-4 can create numerous distinct and reasonable scenarios.  An analyst could use a tool like this to generate many scenarios, pick those that seem most relevant, and then ask the LLM to create initial plans for each of the selected scenarios.

\subsection{Porter's Five Forces}

While the previous subsection provides an example of using AI for strategy generation, we now illustrate how an LLM can conduct strategy evaluation, through a Porter's Five Forces analysis of the business school industry (for detailed prompts and outputs, see Appendix~\ref{app:five-forces}).  Three noteworthy features can be observed in the responses we received.  First, the LLM's answer is comparable in terms of depth and breadth of an MBA student's analysis, covering essential aspects like faculty quality, brand reputation, international partnerships, research funding, accreditations, and rankings.  Second, the LLM provides a detailed assessment of each force and summarizes their intensity as ``Low,'' ``Moderate,'' or ``High.''  Third, the LLM's ability to include considerations not mentioned in the prompt shows that it is not merely echoing the input it was given but using additional knowledge from its training data.

This brief example suggests that in the future, AI could enhance a manager's SDM process by acting as an analyst.  An AI could generate an analysis almost instantly, a task that may take an MBA intern or junior employee days to complete.  As with human analysts, the manager could further prompt the LLM with questions, request revisions, and then use this as a foundation for further analyses.

\subsection{Devil's advocate}

This SDM tool involves assigning someone to review a strategic plan and provide thorough critiques of its main assumptions and arguments \citep{Schw84b}.  It can be used for both strategy generation and evaluation.  To implement the devil's advocate using AI, we provided a paragraph describing the plan of a business school that received one billion dollars from a donor.  The school aims to reposition itself as a business school with a strong focus on technology and entrepreneurship, and a curriculum geared towards future business trends.  In response, the LLM generated 15 contrarian arguments.  For instance, it mentioned the volatile nature of technology and how focusing too much on current trends could lead to rapid obsolescence of the curriculum.  It also mentioned ethical concerns with respect to the school investing in student startups, as well as challenges a disruptive business school might face with accreditation.  All of these are reasonable counter-arguments, which an expert in the field might come up with.

This example suggests that managers could effectively use AI as a devil's advocate.  There are two main reasons for this.  First, traditionally, employing a devil's advocate is resource-intensive, requiring days of work from skilled personnel to develop contrarian arguments.  An LLM reduces these costs, allowing more frequent use.  Second, unlike humans, LLMs do not experience social inhibitions when debating with authority figures, such as a CEO.  This lack of inhibition allows for a more exhaustive and unbiased analysis, enhancing a manager's ability to develop unique and contrarian strategies, crucial for creating value \citep{Feli20}.

\subsection{Wisdom of the crowd}\label{sec:virtual-woc}

The ``wisdom of the crowd'' refers to the use of a large number of individuals to either (1) generate new ideas (a.k.a. crowdsourcing; \citealt{Afua12}) or (2) select among ideas via voting or other aggregation mechanisms \citep{Suro04}.  Rather than developing our own examples about how one could use AI to implement the wisdom of the crowd in the context of strategy, here we build on examples developed in the fields of political science and marketing, that are not difficult to translate to strategy contexts.  \citet{Argy23} introduce the idea of ``silicon sampling''---using LLMs to simulate human samples---and show that these virtual individuals exhibit political and public opinions that are similar to their human counterparts.  They show, for instance, that the election outcomes in a region can be accurately forecast by asking simulated individuals---crafted to reflect the region's sociodemographic makeup---to ``vote'' on political candidates.  Along the same lines, but in the context of social science and market research, \citet{Hort23} and \citet{Bran23} show that LLM-generated responses are consistent with economic theory predictions and exhibit realistic willingness-to-pay estimates.

In the context of strategy, managers could use LLMs to generate virtual crowds and draw upon their knowledge to aid SDM.  For instance, while designing a product, managers could test its appeal with a simulated sample of hundreds of target customers and use their feedback to refine the product's design (see \citealt{Olen23} for an alternative approach at designing a value proposition with the help of an LLM).  If, as in previous research on political science and market research, these virtual crowds are good proxies of their real-world counterparts, using virtual crowds may provide a quick and inexpensive way of testing business hypotheses.  This could enable strategists to accelerate processes like the prototype--test cycle outlined in the lean startup methodology \citep{Ries11}.  It could also help them obtain more robust insights from larger samples than what is typically feasible.

\bigskip

The re-imagined, AI-augmented SDM tools we have described so far offer a glimpse of what AI-augmented SDM may look like in the future.  How effective these AI-augmented SDM tools can be depends on the ability of LLMs to generate and evaluate strategies.  The next section empirically evaluates these abilities.

\section{Evidence on AI-augmented strategy generation and evaluation}\label{sec:evidence}

To measure the ability of LLMs to generate and evaluate strategies, we leverage data from two real-world entrepreneurship contexts.  First, we collaborated with a leading European startup accelerator to experimentally explore how LLMs generate strategies relative to entrepreneurs.  Second, we used data from a startup competition at an elite business school to assess the ability of LLMs to evaluate strategies vis-à-vis experienced venture capital and angel investors.

We find that LLM-generated strategies attract at least as much investor interest as those of entrepreneurs admitted to a leading startup accelerator, and that LLM-evaluations of business plans are positively correlated with those of experienced investors.  This suggests that current AI may have the capacity to generate and evaluate strategies, offering the possibility of a more efficient exploration of strategies than relying solely on human input.

\subsection{AI and strategy generation}\label{sec:generation}

We sourced entrepreneurial business plans submitted to a European startup accelerator program that provided 30,000 euros of funding, mentorship, office space, marketing and legal services to each accepted startup.  We received all submitted applications from 2021--2022, which consisted of 309 business plans.  We manually filtered out any plans that were incomplete or written in the local language, and randomly selected ten business plans (five accepted and five rejected) from this remaining set (see Appendix~\ref{app:generation} for details).

For each business plan, we created an AI-generated version using GPT-3.5.\footnote{All the analyses in this section use GPT-3.5, as this was the only LLM at the time that offered a sufficiently large context window (16,000 tokens) that could fit the long prompts and responses needed to conduct the empirical studies.}  Business plans submitted to the program answered four main questions, describing the ``Problem,'' ``Solution,'' ``Market,'' and ``Go-to-market strategy'' of the startup.  We prompted the LLM using the entrepreneur's original response to the ``Problem'' section and asked it to generate the remaining business plan, to obtain an LLM-generated strategy to solve the problem described (see full prompts in Appendix~\ref{app:generation}).

To compare LLM-generated strategies to those developed by entrepreneurs, we designed and ran an experiment across 250 evaluators, recruited online to help evaluate business plans for the startup accelerator.  Evaluators were US-based and had experience investing in angel syndicates, venture capital, or private equity funds.  Evaluators had on average 5 years of investment experience, 16 years of work experience overall, and 7 years of managerial experience (see Appendix Table~\ref{tbl:sumstats} for summary statistics on the full set of demographic attributes).  The experiment was pre-registered at the AEA RCT Registry and analyzed without deviation.\footnote{Csaszar, F., H. Ketkar and H. Kim.  2024.  ``AI and strategic decision-making (generation experiment).''  AEA RCT Registry.  February 16.  \url{https://doi.org/10.1257/rct.11942-3.0}}

Each evaluator was sent a survey link that provided ten business plans to assess.  For each business plan, each evaluator was randomly assigned to one version (i.e., the entrepreneur- or the LLM-generated version).  This was a within-subject design, meaning that that every evaluator saw at least one entrepreneur- and LLM-generated business plan across the ten plans they assessed, but never saw both versions for each business.\footnote{This differs from across-subject experimental designs that randomly assign participants (e.g., evaluators or firms) to receive AI-driven treatments, such as in \citet{Otis23} and \citet{Dell23}.  In within-subject experimental designs, participants essentially serve as their own control groups by providing baseline measures to compare with different treatment conditions.  For example, in this experiment, we compare the scores assigned to LLM-generated plans relative to entrepreneur-generated plans, controlling for evaluator fixed effects.} Moreover, the plans were not labeled as entrepreneur- or LLM-generated, so that the evaluator could not differentially evaluate plans based on this knowledge.  All evaluations were requested and completed in February 2024.

Each evaluator assessed a set of ten business plans displayed in randomized order.  They were asked to evaluate each business plan on a scale of 1 (low) to 10 (high) on its key attributes---innovation and value proposition, execution plan, potential to invest in the idea, viability, and writing quality\footnote{These questions were adapted from scoring rubrics used by accelerators and startup competitions.}---which were consolidated into a standardized sum.  The evaluators were also asked to indicate: (1) whether they recommended accepting this startup into the accelerator program (yes/no); (2) whether they were interested in being introduced to the startup (yes/no); (3) how likely they were to invest in the startup (a range from 0 to 100), building on measures used in \citet{Huan15}, \citet{Bern17}, and \citet{Bapn19} to assess entrepreneurial venture quality.  Appendix Figure~\ref{fig:rubric} displays the full rubric that evaluators were asked to use for their evaluations.

To compare entrepreneur-generated business plans with those generated by the LLM, we analyzed the following econometric specification for all outcomes $y_{ipj}$ for each version $i$ of the plan $p$ evaluated by evaluator $j$:
\begin{equation}\label{eq:specification}
y_{ipj} = \beta \mathrm{AI}_{ipj} + \gamma_p + \delta_j + \varepsilon_{ipj},
\end{equation}
where $\mathrm{AI}_{ipj}$ is a binary indicator that takes value 1 for business plan versions generated by the LLM and 0 otherwise; $\gamma_p$ controls for business plan fixed effects, $\delta_j$ controls for evaluator fixed effects, and $\varepsilon_{ipj}$ is an idiosyncratic error term.  The coefficient $\beta$ identifies the difference in outcome variables for LLM-generated versions of the business plan relative to the original entrepreneurs' versions, and is the main coefficient of interest.  This is estimated with robust standard errors at the business plan level, which is the unit of randomization \citep{Abad23}.

Table~\ref{tbl:gen} indicates that, on average, plans created by LLMs were rated higher by 0.14 standard deviations ($p<0.001$).  In nearly all evaluated aspects---such as the execution plan, the investment potential of the idea, and the business's viability---LLM-generated plans consistently scored between 0.07 and 0.23 standard deviations higher, as detailed in Appendix Table~\ref{tbl:factors}.  Evaluators were also 5 percentage points more likely to recommend accepting LLM-generated plans to the accelerator ($p=0.003$), 3 percentage points more likely to be interested in being introduced ($p=0.094$), and 3 percentage points more likely to invest in these businesses ($p=0.007$) compared to the original plans crafted by entrepreneurs.

\begin{table}
\begin{center}
\caption{Comparison of entrepreneur- vs.\ LLM-generated business plans.}\label{tbl:gen}
\def\sym#1{\ifmmode^{#1}\else\(^{#1}\)\fi}
\begin{tabular}{l*{4}{c}}
\hline\hline
                    &\multicolumn{1}{c}{(1)}&\multicolumn{1}{c}{(2)}&\multicolumn{1}{c}{(3)}&\multicolumn{1}{c}{(4)}\\
                    &\multicolumn{1}{c}{Evaluation}&\multicolumn{1}{c}{Acceptance}&\multicolumn{1}{c}{Interest in}&\multicolumn{1}{c}{Investment}\\
                    &\multicolumn{1}{c}{Index}&\multicolumn{1}{c}{}&\multicolumn{1}{c}{Introduction}&\multicolumn{1}{c}{Likelihood}\\
\hline
LLM                 &        0.14\sym{***}&        0.05\sym{***}&        0.03\sym{*}  &        2.55\sym{***}\\
                    &      (0.03)         &      (0.02)         &      (0.02)         &      (0.94)         \\
Plan FE             &         Yes         &         Yes         &         Yes         &         Yes         \\
Investor FE         &         Yes         &         Yes         &         Yes         &         Yes         \\
\hline
Observations        &        2500         &        2500         &        2500         &        2500         \\
Mean (entrepreneur)     &       -0.07         &        0.57         &        0.51         &       46.69         \\
SD (entrepreneur)       &        1.04         &        0.50         &        0.50         &       30.41         \\
\hline\hline
\end{tabular}

\end{center}
{\footnotesize \emph{Notes.} \textit{Evaluation Index} is an index variable of business plan evaluation scores, standardized.  \textit{Acceptance} is a binary variable indicating whether the evaluator recommended accepting the plan.  \textit{Interest in Introduction} is a binary variable indicating evaluator interest in being introduced to the business.  \textit{Investment Likelihood} indicates evaluators' stated likelihood of investing in the startup and ranges from 0 to 100.  Standard errors are reported in parentheses.  ${}^{***} p<0.01$, ${}^{**} p<0.05$, ${}^{*} p<0.1$.}
\end{table}

\begin{figure}
\begin{center}
\caption{Comparison of entrepreneur- vs.\ LLM-generated business plans.}\label{fig:gen}
\includegraphics[width=\linewidth]{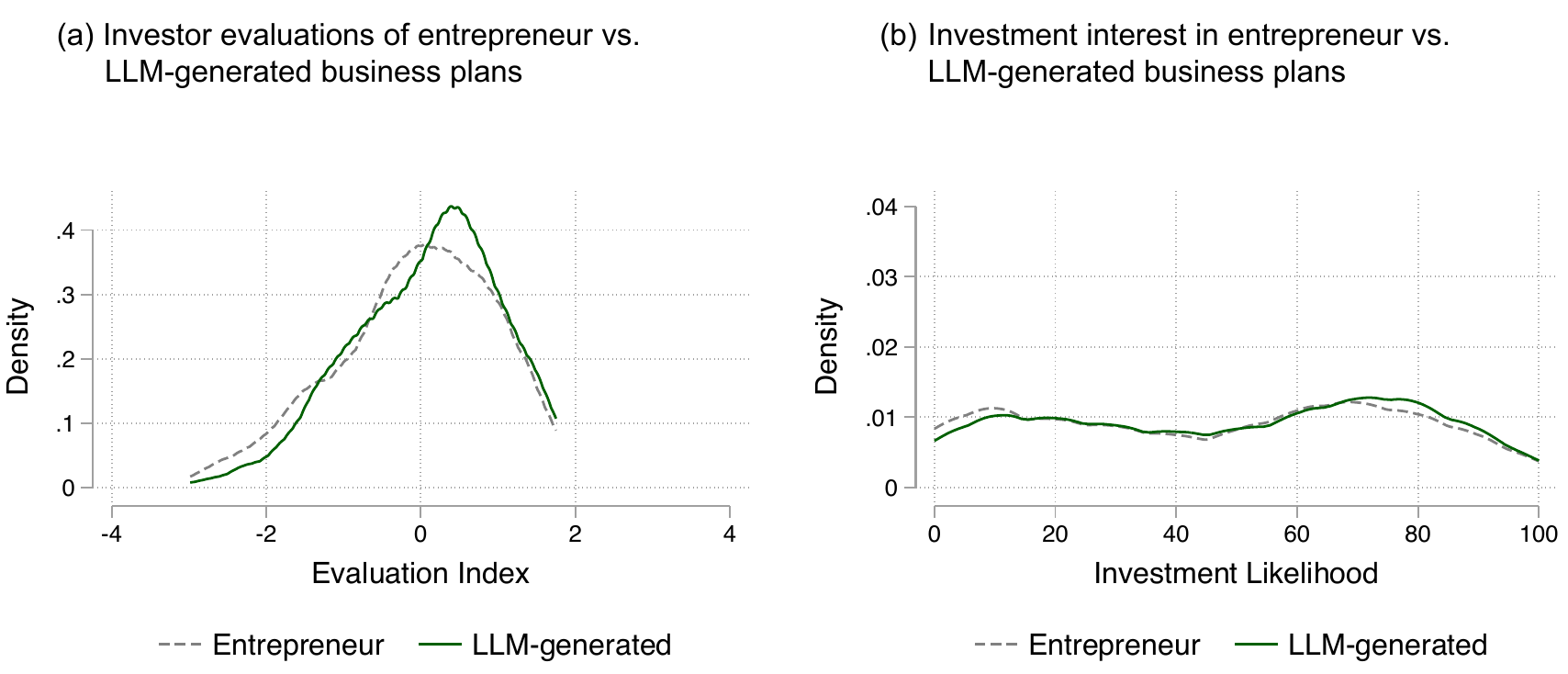}
\end{center}
{\footnotesize \emph{Notes.}  Panel (a) shows kernel density plots of evaluators' scores by plan type, standardized into $z$-scores.  Panel (b) shows kernel density plots of investors' stated likelihood of investing in the business by plan type.  Kolmogorov-Smirnov test \textit{p-value}: (a) 0.002; (b) 0.09.}
\end{figure}

We observe this pattern across the full distribution of plans.  Figure~\ref{fig:gen} plots the distributions of scores and investment likelihoods across both plan types, showing that those of LLM-generated business plans are shifted slightly to the right of entrepreneurs' plans.  This suggests that strategies generated by LLMs receive higher scores and more investment interest across the full distribution, and especially in the left tail of the distribution (i.e., lower-rated plans).  The differences in these distributions are statistically significant ($p$-values of 0.002 and 0.09 for panels (a) and (b), respectively, as determined by the Kolmogorov--Smirnov test).

One potential concern that these results raise is that scores for LLM-generated plans may be inflated by the LLM versions being better written.  To reduce this possibility, the experiment compared LLM-generated plans against entrepreneurs' original business plans edited by an LLM to remove any grammar and spelling mistakes (see Appendix~\ref{app:generation} for details).  Our results thus reflect that holding writing errors constant, the LLM-generated plans on average received similar scores and investment interest compared to business plans developed by entrepreneurs applying to a leading accelerator.  Consistent with this, Appendix Table~\ref{tbl:factors} shows that the LLM-generated plans received higher scores across many dimensions other than writing quality relative to entrepreneurs (columns 2--5).

Another concern may be that evaluator scores do not provide informative signals.  While this is certainly possible, evaluator scores appear to be correlated with independent decisions taken by the startup accelerator, which have real-world implications for startups in practice.  Evaluators in the experiment rated accepted entrepreneurs' business plans more highly than those that were rejected, by approximately 0.6 standard deviations (Table~\ref{tbl:het-accepted}).  They were also 15 percentage points more likely to accept accepted entrepreneurs' business plans, 19 percentage points more likely to be interested in being introduced to them, and 12 percentage points more likely to invest in these businesses---providing some validation that their scores contained a signal correlated with real-world outcomes.

\begin{table}
\begin{center}
\caption{Comparison by whether the entrepreneur's plan was accepted by the accelerator.}\label{tbl:het-accepted}
\def\sym#1{\ifmmode^{#1}\else\(^{#1}\)\fi}
\begin{tabular}{l*{4}{c}}
\hline\hline
                    &\multicolumn{1}{c}{(1)}&\multicolumn{1}{c}{(2)}&\multicolumn{1}{c}{(3)}&\multicolumn{1}{c}{(4)}\\
                    &\multicolumn{1}{c}{Evaluation}&\multicolumn{1}{c}{Acceptance}&\multicolumn{1}{c}{Interest in}&\multicolumn{1}{c}{Investment}\\
                    &\multicolumn{1}{c}{Index}&\multicolumn{1}{c}{}&\multicolumn{1}{c}{Introduction}&\multicolumn{1}{c}{Likelihood}\\
\hline
LLM                 &        0.29\sym{***}&        0.07\sym{***}&        0.08\sym{***}&        5.88\sym{***}\\
                    &      (0.04)         &      (0.03)         &      (0.02)         &      (1.32)         \\
LLM * Accepted Application&       -0.29\sym{***}&       -0.04         &       -0.10\sym{***}&       -6.66\sym{***}\\
                    &      (0.06)         &      (0.04)         &      (0.04)         &      (2.01)         \\
Accepted Application&        0.55\sym{***}&        0.15\sym{***}&        0.19\sym{***}&       11.84\sym{***}\\
                    &      (0.07)         &      (0.04)         &      (0.04)         &      (2.20)         \\
Plan FE             &         Yes         &         Yes         &         Yes         &         Yes         \\
Investor FE         &         Yes         &         Yes         &         Yes         &         Yes         \\
\hline
Observations        &        2500         &        2500         &        2500         &        2500         \\
\hline\hline
\end{tabular}

\end{center}
{\footnotesize \emph{Notes.} \textit{Evaluation Index} indicates an index variable of business plan evaluation scores, standardized.  \textit{Acceptance} is a binary variable indicating whether the evaluator recommended accepting the plan.  \textit{Interest in Introduction} is a binary variable indicating evaluator interest in being introduced to the business.  \textit{Investment Likelihood} indicates evaluators' stated likelihood of investing in the startup and ranges from 0 to 100.  Robust standard errors are reported in parentheses.  ${}^{***} p<0.01$, ${}^{**} p<0.05$, ${}^{*} p<0.1$.}
\end{table}

We find that LLM-generated plans especially outperformed plans that were rejected by the startup accelerator (Table~\ref{tbl:het-accepted}).  LLM versions were 7 percentage points more likely to be recommended for acceptance ($p=0.003$), 8 percentage points more likely to garner interest in being introduced ($p=0.001$), and 6 percentage points more likely to be invested in ($p<0.001$) compared to entrepreneurs' plans that were ultimately rejected by the accelerator, but these effects appear to be canceled out for plans that were accepted into the accelerator---suggesting that the LLM may be most helpful in generating strategies for entrepreneurs in the lower tail of the distribution, consistent with findings in other contexts on worker productivity \citep{Bryn23,Noy23}.

Together, these results raise the possibility that AI may be used to help entrepreneurs in generating strategies as they start a venture or improve their business model, enabling them to consider a potentially broader and higher-quality set of strategies.

\subsection{AI and strategy evaluation}\label{sec:evaluation}

To explore how AI evaluates strategies, we partnered with a startup competition at an elite business school that provided seed funding and prizes amounting to 85,000 euros to the winning startups.  Business plans submitted to the first round of the competition were evaluated by a panel of judges composed of venture capitalists and angel investors, many of whom had also been entrepreneurs themselves.

We received all 731 business plans submitted to the last ten cycles of the competition, each about 1,500 words long.  We identified 138 plans that were described fully using text rather than images, which represented 19\% of all plans submitted during this period.  Each plan had been evaluated by three to five judges assigned to the plan based on their expertise and experience.  This provided us with a dataset of 541 evaluations by 137 unique judges for the 138 business plans.

We then had the LLM evaluate these business plans using a series of prompts (fully detailed in Appendix~\ref{app:evaluation}).  We instructed the LLM to evaluate the business plans using the same rubric employed by the competition judges, which involved assigning scores ranging from 0\% to 100\% across the following categories: (1) team, (2) execution plan, (3) financials, (4) writing, (5) innovativeness of the value proposition, (6) market opportunity, (7) competitive advantage of the business, (8) extent to which the product/service addresses actual customer needs, (9) return on investment potential, and (10) viability of the business in the long run.  We generated three iterations of LLM scores and averaged across them.  This process provided us with LLM scores on the same dimensions and scale as human judges, across which we also took an average.  We standardized both sets into $z$-scores.

\begin{figure}
\caption{Investor vs.\ LLM evaluations of business plans.}\label{fig:human-vs-ai-scoring-eval}
\includegraphics[width=\linewidth]{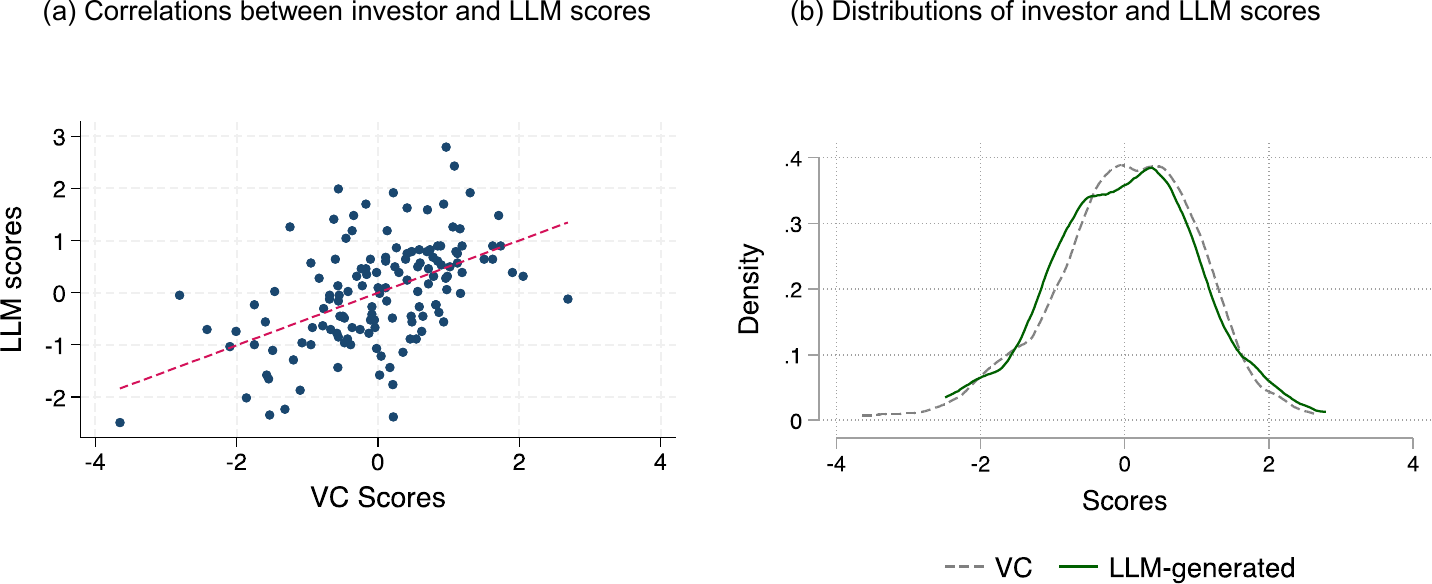}
{\footnotesize \emph{Notes.}  Panel (a) shows a scatterplot of standardized LLM-generated evaluation scores versus standardized investor scores of 138 business plans submitted to a startup competition.  The correlation coefficient is 0.52.  Panel (b) shows kernel density plots of the investor and LLM-generated evaluation scores.  The Kolmogorov--Smirnov test $p$-value is 0.86.}
\end{figure}

Figure~\ref{fig:human-vs-ai-scoring-eval}(a) shows that the LLM evaluations and those of judges (labeled ``VC scores'') are positively correlated, with an average correlation of 0.52.\footnote{We also computed Inter-Rater Reliability (IRR) scores to compare the agreement between the average LLM and VC scores, and between the individual VC scores, using the Intraclass Correlation Coefficient (ICC).  The results show that the IRR between AI--VC ($\text{ICC}=0.51$) is higher than between individual VCs ($\text{ICC}=0.25$), indicating greater consistency between the LLM and VC scores.  We further compared these measures using Fisher's $z$-scores, which yielded 0.56 for AI--VC and 0.25 for individual VCs, a difference that is statistically significant at the 10\% level.}  This correlation is robust across business plans submitted within the LLM's training window, as well as afterward (see Figure~\ref{fig:scatterplot-pre-post-training} in Appendix~\ref{app:evaluation}).  Standardized LLM and judges' scores also share similar distributions (Figure~\ref{fig:human-vs-ai-scoring-eval}(b)).

Table \ref{tbl:reg-ai-on-human-gen} shows that across each indicator, the magnitude of the correlation between the two scores vary, with the scores correlating more highly on the evaluations of the team, the execution plan, the financials, and the competitive advantage of the business.  The $R$-squared values suggest that overall LLM scores explain approximately 29\% of the variation in the investor scores.

\begin{sidewaystable}
\caption{Comparison of investor- vs.\ LLM-generated scores.}\label{tbl:reg-ai-on-human-gen}
\adjustbox{max width=\textwidth}{\setlength{\tabcolsep}{0.5\tabcolsep}
\def\sym#1{\ifmmode^{#1}\else\(^{#1}\)\fi}
\begin{tabular}{l*{11}{c}}
\hline\hline
                    &\multicolumn{1}{c}{(1)}&\multicolumn{1}{c}{(2)}&\multicolumn{1}{c}{(3)}&\multicolumn{1}{c}{(4)}&\multicolumn{1}{c}{(5)}&\multicolumn{1}{c}{(6)}&\multicolumn{1}{c}{(7)}&\multicolumn{1}{c}{(8)}&\multicolumn{1}{c}{(9)}&\multicolumn{1}{c}{(10)}&\multicolumn{1}{c}{(11)}\\
                    &\multicolumn{1}{c}{VC}&\multicolumn{1}{c}{VC}&\multicolumn{1}{c}{VC}&\multicolumn{1}{c}{VC}&\multicolumn{1}{c}{VC}&\multicolumn{1}{c}{VC}&\multicolumn{1}{c}{VC}&\multicolumn{1}{c}{VC}&\multicolumn{1}{c}{VC}&\multicolumn{1}{c}{VC}&\multicolumn{1}{c}{VC}\\
                    &\multicolumn{1}{c}{Scores}&\multicolumn{1}{c}{Team}&\multicolumn{1}{c}{Execution}&\multicolumn{1}{c}{Financials}&\multicolumn{1}{c}{Writing}&\multicolumn{1}{c}{Innovation}&\multicolumn{1}{c}{ Market}&\multicolumn{1}{c}{ Competitive}&\multicolumn{1}{c}{Product}&\multicolumn{1}{c}{ Investment}&\multicolumn{1}{c}{Viability}\\
                    &\multicolumn{1}{c}{}&\multicolumn{1}{c}{}&\multicolumn{1}{c}{}&\multicolumn{1}{c}{}&\multicolumn{1}{c}{}&\multicolumn{1}{c}{}&\multicolumn{1}{c}{ Opportunity}&\multicolumn{1}{c}{Advantage}&\multicolumn{1}{c}{}&\multicolumn{1}{c}{Potential}&\multicolumn{1}{c}{}\\
\hline
LLM Scores     &        0.52\sym{***}&                     &                     &                     &                     &                     &                     &                     &                     &                     &                     \\
                    &      (0.08)         &                     &                     &                     &                     &                     &                     &                     &                     &                     &                     \\
LLM  Team     &                     &        0.46\sym{***}&                     &                     &                     &                     &                     &                     &                     &                     &                     \\
                    &                     &      (0.07)         &                     &                     &                     &                     &                     &                     &                     &                     &                     \\
LLM  Execution&                     &                     &        0.40\sym{***}&                     &                     &                     &                     &                     &                     &                     &                     \\
                    &                     &                     &      (0.08)         &                     &                     &                     &                     &                     &                     &                     &                     \\
LLM  Financials&                     &                     &                     &        0.57\sym{***}&                     &                     &                     &                     &                     &                     &                     \\
                    &                     &                     &                     &      (0.09)         &                     &                     &                     &                     &                     &                     &                     \\
LLM  Writing  &                     &                     &                     &                     &        0.30\sym{***}&                     &                     &                     &                     &                     &                     \\
                    &                     &                     &                     &                     &      (0.09)         &                     &                     &                     &                     &                     &                     \\
LLM  Innovation&                     &                     &                     &                     &                     &        0.21\sym{**} &                     &                     &                     &                     &                     \\
                    &                     &                     &                     &                     &                     &      (0.08)         &                     &                     &                     &                     &                     \\
LLM  Market Opportunity&                     &                     &                     &                     &                     &                     &        0.28\sym{***}&                     &                     &                     &                     \\
                    &                     &                     &                     &                     &                     &                     &      (0.08)         &                     &                     &                     &                     \\
LLM  Competitive Advantage&                     &                     &                     &                     &                     &                     &                     &        0.38\sym{***}&                     &                     &                     \\
                    &                     &                     &                     &                     &                     &                     &                     &      (0.09)         &                     &                     &                     \\
LLM  Product  &                     &                     &                     &                     &                     &                     &                     &                     &        0.24\sym{***}&                     &                     \\
                    &                     &                     &                     &                     &                     &                     &                     &                     &      (0.07)         &                     &                     \\
LLM  Investment Potential&                     &                     &                     &                     &                     &                     &                     &                     &                     &        0.25\sym{***}&                     \\
                    &                     &                     &                     &                     &                     &                     &                     &                     &                     &      (0.09)         &                     \\
LLM  Viability&                     &                     &                     &                     &                     &                     &                     &                     &                     &                     &        0.35\sym{***}\\
                    &                     &                     &                     &                     &                     &                     &                     &                     &                     &                     &      (0.09)         \\
Constant            &       0.02         &        0.28         &        0.03         &       0.21         &       0.28         &        0.34         &        0.16         &       0.09         &        0.25         &       0.17         &       0.02         \\
                    &      (0.25)         &      (0.21)         &      (0.19)         &      (0.28)         &      (0.31)         &      (0.27)         &      (0.22)         &      (0.24)         &      (0.23)         &      (0.32)         &      (0.27)         \\
Round FE            &         Yes         &         Yes         &         Yes         &         Yes         &         Yes         &         Yes         &         Yes         &         Yes         &         Yes         &         Yes         &         Yes         \\
\hline
Observation         &      138.00         &      138.00         &      138.00         &      138.00         &      138.00         &      138.00         &      138.00         &      138.00         &      138.00         &      138.00         &      138.00         \\
Rsquared           &        0.29         &        0.25         &        0.19         &        0.31         &        0.14         &        0.11         &        0.15         &        0.24         &        0.10         &        0.10         &        0.16         \\
Adjusted Rsquared  &        0.24         &        0.19         &        0.12         &        0.26         &        0.07         &        0.04         &        0.08         &        0.18         &        0.03         &        0.03         &        0.10         \\
\hline\hline
\end{tabular}
}
{\footnotesize \emph{Notes.}  This table regresses standardized investor scores on the LLM-generated scores by evaluation component, controlling for fixed effects for each round of the competition.  Robust standard errors are reported in parentheses.  ${}^{***} p<0.01$, ${}^{**} p<0.05$, ${}^{*} p<0.1$.}
\end{sidewaystable}

One possible implication of this correlation is that investors could use AI to screen ideas at scale (e.g., the first round of a startup competition).  Another practical implication is that entrepreneurs could use the feedback from an AI to improve their business plan in the early stages.  For example, in our exercise, the LLM's evaluations included not just scores, but specific feedback about what had been done well or poorly, similar to the feedback a mentor could give an entrepreneur.  Overall, these results suggest that high-quality appraisal and advice---historically available only from a relatively small group of very busy people---may potentially no longer be a bottleneck.

\section{Implications of AI-augmented SDM}\label{sec:implications}

So far we have illustrated how AI-augmented SDM could look like (Section~\ref{sec:how}) and shown that current LLMs possess the key abilities required for SDM---generating and evaluating strategies---at a level comparable to humans participating in leading startup accelerators and business plan competitions (Section~\ref{sec:evidence}).  Overall, our vignettes and empirical studies suggest that LLMs have the potential to reshape the SDM process.  The likelihood of this transformation can only increase as AI continues to advance.

In this section, we try to better understand the implications of this transformation by examining how key strategy concepts may be influenced by the progressive use of AI in strategy.  This is a significant transformation, as SDM has until now been the sole domain of humans.  Of course, because we only have access to the preliminary AI-augmented SDM tools and results presented in sections \ref{sec:how} and \ref{sec:evidence}, what we say should be read as conjectures.  Still, exploring these ideas is worthwhile as they point out many practical challenges and research areas that need attention as AI assumes a more prominent role in strategy.

One important consideration when examining the potential impact of AI on SDM is that AI modifies a fundamental constraint of SDM: bounded rationality \citep{Simo55}.  Specifically, AI can extend the limits of rationality by facilitating the processing of more information, more rapidly, and at a lower cost than what human strategists can achieve.  Since bounded rationality is central to all aspects of the SDM process \citep{Levi11}, relaxing this constraint can produce wide-ranging effects.

Given this multitude of potential effects, it is beneficial to have a systematic way of exploring them.  Here, we leverage Csaszar and Steinberger's \citeyearpar{Csas22} classification of the cognitive processes within firms that are affected by AI: search, representation, and aggregation.  These processes encompass distinct ways by which firms can produce intelligent action \citep{Ocas20}, and have been extensively researched within the information processing literature that has developed within strategy and organizations.  \emph{Search} processes involve finding potential solutions to a problem and trying them out (a classic work in this vein is \citealt{Levi97}; for surveys, see \citealt{Pose18} and \citealt{Baum19}).  \emph{Representation} processes entail developing a model of the problem faced by the firm and using this model to devise an appropriate solution \citep{Huff90,Trip00,Csas16}.  Lastly, \emph{aggregation} processes work by combining a diverse set of weak, preliminary solutions to produce a stronger solution \citep{Csas13b,Csas18}.  Below, we examine how AI could affect these three processes.  For clarity, Table~\ref{tbl:implications-outline} outlines the implications discussed in this section.

\begin{table}\centering
\caption{Outline of proposed implications of AI on search, representation, and aggregation.}\label{tbl:implications-outline}
\renewcommand{\arraystretch}{1.5}
\begin{tabular}{p{1.8in}p{1.8in}p{1.8in}}
\hline
\multicolumn{1}{c}{\textbf{Search}} & \multicolumn{1}{c}{\textbf{Representation}} & \multicolumn{1}{c}{\textbf{Aggregation}} \\
\hline
\begin{tabitem}
  \item Speed
  \item Fitness
  \item Heterogeneity
  \item Competition
  \item Experimentation
\end{tabitem}
&
\begin{tabitem}
  \item Representational complexity
  \item New representations and frameworks
  \item Changing and re-imagining representations
\end{tabitem}
&
\begin{tabitem}
  \item Simulating markets, case discussions, and competitors
  \item Decreasing biases
  \item History-friendly models
  \item Boundary objects
\end{tabitem} \\
\hline
\end{tabular}

\end{table}

\subsection{Search}\label{sec:search}

Solving a problem using search requires trying out alternatives until one good enough (a.k.a. ``satisficing''; \citealt{Simo55}) is found.  Strategy research typically conceptualizes search in terms of traversing a landscape to find a high peak \citep{Levi97}.  Here, we explore how AI might impact search by analyzing its potential effects on the primary outcomes studied in this literature---speed, fitness, and heterogeneity---and how these changes could influence competition and experimentation.

Speed in generating and evaluating alternatives can be significantly increased by AI, as suggested by the studies developed in sections~\ref{sec:generation} and \ref{sec:evaluation}.  For example, a company planning to launch a new product could use AI to generate and evaluate hundreds of potential product ideas.  Managers could then select from the AI-recommended top ideas.  This system would significantly speed up the exploration of different options.  Without AI, creating and analyzing such a vast array of ideas would likely be excessively expensive and time-consuming.  However, with AI, managers could undertake such tasks more efficiently, allowing companies to consider more ideas and potentially launch more new products.  At the population level, faster search could enable a faster pace of competitive evolution or ``clockspeed'' \citep{Fine98}.

Besides speed, the search literature has focused on two other outcomes: fitness (decision-making quality) and heterogeneity (solution variation across firms).  The impact of AI on these outcomes is less clear than on speed, as much depends on the level of sophistication AI can achieve.  We illustrate this point by considering two scenarios: one with current-level AI abilities, termed ``weak AI,'' and another with more advanced AI that could come in the future, termed ``strong AI,'' which surpasses current limitations in terms of reasoning skills and available knowledge.

With weak AI, idea quality and innovativeness would be comparable to human-generated proposals in business plan competitions, but unlikely to be groundbreaking.  Ideas would likely cluster around areas well-represented in LLM training sets.  While search speed would increase, fitness and heterogeneity of solutions may not significantly improve, though they could enhance average fitness in some contexts, especially where knowledge acquisition faces friction \citep{Giro23,Dell23,Otis23}.

Weak AI's limitations stem from LLMs' lack of advanced reasoning capabilities and training data biases \citep{Huan23}.  Without substantial progress, AI may only improve search speed while potentially harming fitness and heterogeneity, as firms might cluster around familiar but mediocre landscape areas \citep{Gave12}.

Strong AI, however, could increase both search speed and fitness levels, potentially discovering opportunities overlooked by even the best human managers.  Firms using strong AI would be more likely to reach all economically viable positions on the landscape.  Heterogeneity across firms would be higher than with weak AI but not necessarily higher than with human managers, as bounded rationality can lead to unprofitable positions that strong AI would avoid.

Note that fitness is not the same as profitability.  With strong AI, firms would find better solutions (e.g., products that create more value or are cheaper to produce).  But that does not mean that firms would be more profitable.  The effect of AI on profitability will also depend on competition.  We come back to this topic in Section~\ref{sec:perf} after having examined the other ways in which AI affect firms' cognitive processes.

Future research should explore situations where AI is most useful for strategic search, considering factors like AI training data, industry-wide AI adoption, and the importance of speed in different sectors.  Since firms often use experimentation to search \citep{Camu20}, examining the role of AI in this process is particularly relevant.  Future strategists could potentially use AI to devise and run experiments ``in silico,'' predicting outcomes using methods like a virtual wisdom of the crowd (Section~\ref{sec:virtual-woc}).  This shift from ``online'' to ``offline'' search \citep{Gave00} opens up the possibility of strategic experimentation without human involvement.  Determining when and to what extent this approach would be effective is an important research question.

\subsection{Representation}\label{sec:representation}

Building on classic research on cognition \citep[\p 61]{Crai43}, strategic representations are conceptualized as simple models that allow for making strategic decisions \citep{Csas18b}.  When these representations are held in the mind, they are referred to as internal or mental representations; meanwhile, when they are embodied in an external artifact, such as a piece of paper, they are referred to as external representations.  Unlike the search approach, which requires trying various alternatives until finding a satisficing one, solving problems with representations involves using a given representation to answer a question directly.  For example, a manager using Porter's Five Forces model can easily address industry-related questions.  For this reason, \citet[\p 132]{Simo96} says ``solving a problem simply means representing it so as to make the solution transparent.''  We explore how AI might impact the complexity, creation, and change of strategic representations.

Strategy typically employs simple representations.  \citet{Levi11} argues that external representations like ``two-by-twos'' in strategy textbooks help managers navigate complex environments.  Similarly, \citet{Bing11} demonstrate that managers use simple heuristics, considering only a few aspects of a problem when making decisions.  \citet{Csas20b} show that the optimal ``representational complexity'' increases as firms can use more information to make decisions.  Consequently, AI's ability to access vast information---current LLMs draw from trillions of words across millions of sources \citep{Borg21}---suggests the complexity of the representations used by AIs will likely surpass that of humans.

The increased representational complexity afforded by AI-augmented SDM may allow for the creation of completely new strategic representations.  So far, the representations in use in strategy have been developed with the bounded rationality of managers in mind.  For instance, Porter's Five Forces likely includes only five forces because it would be unwieldy for human managers to deal with many more factors.  But if humans cease to be the primary users of these representations, this opens the possibility of devising new, more complex representations that can use the additional complexity to make more accurate predictions.  For instance, a future strategy framework could be developed by an AI by extracting all concepts from every strategy paper ever written, analyzing every existing analyst report using these concepts, and then identifying those concepts that prove to be predictive of future firm performance, contingent on firm and industry characteristics.  Given sufficient data, this framework would likely outperform current ones.  Future research could explore representations unconstrained by human bounded rationality and their potential improvements over simple ones.  Using Porter's \citeyearpar[\p 97]{Port91} distinction between frameworks and models, it is possible that future ``AI strategy frameworks'' will combine the breadth of frameworks with the precision of models.

Apart from increasing representational complexity and allowing for the creation of new strategic representations, AI may also affect managers' ability to change representations.  Traditionally, changing a representation involves convincing managers about a new way to understand the world, which is notoriously difficult, time-consuming, and unlikely to happen \citep{Doug92}.  With AI, however, changing a representation could be as simple as changing the prompt of an LLM.  This improved ability to change representations is important, as \citet{Csas16} show that much value can be unlocked by searching over representations.  \citet{Gave16} illustrate the value of changing representations by recounting the story of Merrill Lynch re-imagining how a bank would look if it was more like a supermarket.  Using an LLM, this re-imagining could be repeated almost costlessly for any firm using any target (e.g., re-imagining a bank like a spa, or a restaurant like a theme park).  The AI could create vivid prose, images, and even videos \citep{Broo24} to illustrate how this re-imagined business would look like, which could increase the chances that human managers can understand and correctly assess the value of the proposal.  Moreover, using a method similar to the one developed in Section~\ref{sec:evaluation}, part of the evaluation of these different reconceptualizations could be automated, allowing managers to focus on the best ideas.

\subsection{Aggregation}

Due to the simplified nature of representations, predictions made using them are inherently fallible.  Aggregation aims to mitigate this by combining predictions from diverse sources.  Aggregation is important in SDM, as strategic decisions are rarely made by a single person \citep{Csas13b}.  Two approaches to aggregation are ensembling and specialization.  In ensembling, a question is broadcast to multiple decision-makers, whose responses are aggregated to improve decision quality (e.g., through majority voting).  In specialization, decision-makers tackle specific parts of a problem, and aggregation involves integrating these parts into a coherent decision (this depends on factors such as how the problem is divided, who reports to whom, and who has decision-making authority).  The primary research question in the aggregation literature is: What is the effect of different aggregation structures on decision quality?  Key variables include task characteristics (e.g., generation versus evaluation), environmental factors (e.g., complexity and uncertainty), and individual attributes (e.g., expertise and incentives).  We explore how AI might influence our understanding of aggregation, in both ensembling and specialization contexts.

Ensembling, also known as the ``wisdom of the crowd,'' is used in SDM for generating ideas through crowdsourcing \citep{Afua12}, as well as for evaluating ideas via opinion polls and committee voting \citep{Csas12}.  The direct effect of AI on these approaches is that, as outlined in Section~\ref{sec:virtual-woc}, it enables the formation of ``virtual crowds'' representing different expertise and demographic characteristics.  Strategists could use a virtual crowd to simulate a market to ``test'' a new product's appeal or discuss strategic decisions with virtual advisors.  A manager might conduct a case discussion with an audience of different expert personas, akin to a manager taking center stage in a live case study within a business school classroom.

The evidence we show in Section~\ref{sec:evidence} suggests that virtual crowds could add value to applications like these, as they revolve around generating and evaluating strategic alternatives.  However, more research is needed to understand how to best design and use virtual crowds.  Key questions include how accurately various persona types mimic their real-world counterparts, the ideal mix of personas for specific tasks, the most effective aggregation mechanism for different tasks, how virtual and human ensembles compare, and when human--AI ensembles are beneficial (for initial work in this area, see \citealt{Pura21,Chou23,Dosh24,Rais24,Eick24}).

The other approach to aggregation, specialization, also poses a host of practical opportunities and research questions.  One area of application is in reducing biases in SDM processes, where different managers may push for their own interests rather than what is best for the firm.  For instance, a host of biases have been identified in prior literature, including sugarcoating \citep{Fang14}, evaluation apprehension \citep{Reit15}, self-promotion \citep{Keum17}, and organizational pressure \citep{Piez23}.  Future studies could explore strategies for employing AI to diminish these biases.  One potential strategy could involve presenting AI recommendations prior to those from humans \citep{Sun22}.

Another application is simulating role-playing and decision-making processes.  For example, a CEO might simulate a conversation with a competitor's CEO.  More ambitious simulations could re-enact historical strategic decisions.  The literature on ``history-friendly models'' \citep{Male16} has developed models to study industry evolution post-key innovations.  LLMs could enhance these models by creating virtual personas of managers who made pivotal decisions and using these personas in simulations (see \citealt{Hua24} for a related application in the context of world wars).  Repeated simulations can reveal the likelihood of various outcomes and offer insights for future decisions.

As AI advances, its capabilities may surpass those of human managers in various domains at different rates, potentially replacing certain tasks or management team members.  The interaction between AI systems and human managers will become crucial.  Current research on top management team communication emphasizes the importance of ``boundary objects''---tools like PowerPoint presentations and diagrams that aid communication among diverse individuals \citep{Carl02,Kapl11b}.  With the rise of AI-human interactions within top management teams, a new set of boundary objects and communication protocols will be essential to ensure that no vital information is overlooked, misunderstood, or mistrusted \citep{Lebo22}.

\section{Conclusion and future directions}\label{sec:conclusion}

We started this article by posing three research questions.  Our answers are: (1) AI-augmented SDM is plausible and may initially consist of LLMs applying strategy frameworks and answering questions (as illustrated in sections \ref{sec:how} and \ref{sec:evidence}), similar to an analyst assisting a strategist; (2) Current LLMs can achieve human-comparable performance in realistic strategy tasks involving generation and evaluation, the two main outputs of SDM (as shown in sections \ref{sec:generation} and \ref{sec:evaluation}); and (3) AI will likely have many implications for strategy research, as it relaxes bounded rationality, a fundamental constraint of SDM, altering many underlying processes (as shown in Section \ref{sec:implications} regarding search, representation, and aggregation).

To develop these answers, we created example uses of AI in SDM, aiming to inspire further exploration by others.  We also devised ways to conceptualize AI's effects on SDM by examining its core outputs---generation and evaluation---and cognitive processes---search, representation, and aggregation.  Our conceptualization complements previous research portraying AI as primarily reducing firms' prediction costs \citep{Agra18,Bryn21} and increasing worker productivity \citep{Noy23,Bryn23,Dell23}.

We conclude by examining how research on AI and strategy may continue to develop.  We approach this in two ways: First, we study how the SDM changes we have examined may ultimately affect firm performance, providing a framework that ties together the discussed elements with strategy outcomes.  Second, we explore how our research may influence the theory-based view (TBV), informing a discussion on the potential limits of AI's impact on strategy.  These perspectives offer insights into the future trajectory of AI and strategy research, highlighting both opportunities and boundaries of AI's influence on SDM processes and outcomes.

\subsection{Effects on performance}\label{sec:perf}

So far, our discussion has focused on AI's effect on the SDM process.  However, strategy ultimately concerns firm performance.  In this section, we propose a framework to understand AI's effects on firm performance.

\begin{figure}\centering
\includegraphics[width=\linewidth]{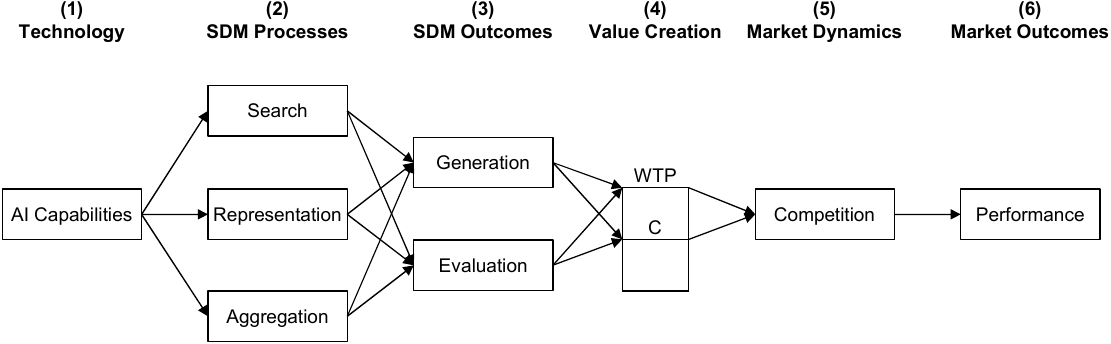}
\caption{Framework connecting the use of AI in SDM to firm performance.}\label{fig:boxes-and-arrows}
\end{figure}

Figure~\ref{fig:boxes-and-arrows} illustrates our framework, integrating the elements discussed so far (columns 1--3: technology, SDM processes, and SDM outcomes) with the customary determinants of firm performance (columns 4--6: value creation, market dynamics, and market outcomes).

Reading the framework from left to right summarizes our theorizing and connects it to firm performance.

The first set of arrows, from technology to SDM processes, corresponds to our discussion in Section 4 about how new AI capabilities may shape search, representation, and aggregation in novel ways.  For example, AI may allow firms to generate numerous alternatives, represent problems with more complexity, or aggregate opinions across a crowd of virtual experts.

The second set of arrows, from SDM processes to SDM outcomes, show how specific SDM processes affect strategy generation and evaluation.  As discussed in Section \ref{sec:search}, AI could generate diverse strategies by searching a larger possibility space, changing representations across different perspectives, or interacting with virtual personas.  Alternatively, AI could aid strategy evaluation by ranking proposed strategies, critiquing them using the devil's advocate approach, or offering opinions from multiple perspectives.

While our paper suggests that AI aids strategy generation and evaluation, its effect on firm performance depends on competitive dynamics.  These dynamics will differ based on AI's technological capabilities.  Similarly to Section \ref{sec:search}, we split the following analysis into ``weak'' and ``strong'' cases.  We also add an intermediate scenario called ``progressing,'' denoting a state of flux between these extremes, in which AI capability on SDM improves (much like it has since the introduction of GPT-3 in November 2022).

Under weak AI (AI plateaus at current levels), firms that adopt it might see increased productivity as AI helps reduce costs.  However, this could also lead to intensified competition due to lower entry barriers, reducing firm performance overall.  Moreover, among firms that adopt AI, differences may diminish as they converge on well-known AI applications, resulting in commoditization.  Over the long term, after a prolonged period of stable weak AI, firms might learn to use AI effectively, making it a standard tool like Excel, with no significant differential impact on profits.  The speed at which this equilibrium is achieved depends on the development of high-quality ``strategy apps'' that democratize AI-augmented SDM for firms of all sizes.\footnote{Academics and consultants may be best positioned to create these.}

Under progressing AI (continuous improvement), firms are likely to vary in their ability to adopt and integrate these new capabilities.  Some firms may develop more advanced strategic decision-making (SDM) processes, enabling them to ``outthink'' competitors.  Recent research highlights significant differences in companies' AI adoption \citep{McEl24} and implementation capabilities \citep{Pura21,Sun22,Agra23}, which become even more relevant in the context of progressing AI.  Moreover, the diffusion of best practices for AI integration is expected to be slow, as observed with previous technological advancements \citep{Bloo12}.  Consequently, firms that effectively adopt and leverage AI may enjoy a series of temporary advantages.  This scenario aligns with Schumpeterian competition, where profits accrue to those better able to innovate \citep[\p 369]{Maho92}.

Under strong AI or ``artificial superintelligence'' (achieving capabilities surpassing the brightest humans; \citealt{Bost14}), firms with complementary assets would have an advantage \citep{Berg23}.  These assets could be AI-specific (e.g., OpenAI's access to advanced GPT versions) or traditional assets that ensure profitability despite increased competition (e.g., Pfizer's patent portfolio, Apple's device ecosystem, or Starbucks' brand and supply-chain relationships).  Competition in this scenario is Ricardian: profits depend on extracting rents from scarce assets \citep[\pp 364--365]{Maho92}.

This analysis suggests that as AI capabilities progress, the sources of competitive advantage will shift, necessitating changes in strategy education and practice.  As AI advances, firms of all sizes could potentially access a ``McKinsey-in-a-box,'' democratizing SDM and potentially increasing the overall sophistication of business strategies across the economy.  Ultimately, market forces may drive AI adoption in SDM, with shareholders demanding that companies leverage AI capabilities to remain competitive, potentially reshaping the role of human strategists.  These developments underscore the importance of understanding AI's integration into SDM processes, while also highlighting the need to understand its potential limitations, which we do next.

\subsection{AI and the TBV}

The TBV describes SDM as a three-stage process: forming testable theories, experimenting to test these theories, and updating beliefs based on results \citep[\p 385]{Zell23}.  Juxtaposing the findings and ideas discussed in our paper with those of the TBV reveals some frictions.  Examining these frictions can provide insights into the potential limitations and opportunities of AI in SDM, helping refine our understanding of how AI can be integrated into strategy formation.  As the use of AI in SDM is nascent and rapidly evolving, time will reveal how these frictions are resolved.

We explore the potential impacts of AI on each stage of the TBV, presenting arguments for and against its successful use in SDM.  Additionally, we consider whether theories themselves are necessary in AI-driven strategy, which may reshape our understanding of the entire process.

\subsubsection{Building theories with AI}

We have proposed that AI could help generate strategies, potentially enhancing the speed and scale at which firms can develop strategic theories.  AI systems like LLMs can rapidly process vast amounts of information and generate numerous strategic options and select the most likely to be successful, which could accelerate the theory-building process.  Additionally, AI could assist in identifying patterns and connections that humans might overlook \citep{Chou19,Shre21}, potentially leading to novel theories and strategic insights \citep{Kang24,Ludw24}.

However, several arguments in the TBV can be interpreted as questioning AI's capacity to develop valuable strategy theories.  First, AI-generated theories risk merely replicating existing ideas rather than producing truly novel concepts, contradicting TBV's emphasis on originality in strategic thought \citep[\p 131]{Feli09}.  That is, AI may fail to capture the unique, idiosyncratic insights crucial to theory formation \citep[\p 386]{Zell23}.  Second, the TBV posits that strategy involves creating future realities, potentially beyond the scope of current AI systems, which excel at recognizing patterns in existing data \citep{Feli24}.  Finally, the TBV argues that SDM may not fundamentally involve traditional search or information processing, questioning AI's ability to effectively contribute to strategic theory formulation \citep[\p 270]{Feli14}.

\subsubsection{Testing theories with AI}

We have argued that AI could enhance managers' ability to conduct thought experiments and simulations, potentially making theory testing more comprehensive.  AI may improve the accuracy of strategy theory testing in two ways.  First, it can assist in rapidly evaluating and filtering numerous strategic options, something boundedly rational humans find difficult.  Second, AI could be used to test theories within a simulated market and help predict competitors' potential responses.  Thus, AI-powered simulations could mitigate the risk of false positives and negatives when launching new products.

However, some ideas in the TBV could caution against over-reliance on AI for testing theories, as this could lead to a disconnection from real-world market conditions and focus on quantifiable metrics at the expense of hard-to-measure, yet critical qualitative insights \citep[\p 140]{Feli09}.  The TBV underscores that real-world experimentation and feedback are necessary to validate strategic theories \citep[\p 385]{Zell23}.

\subsubsection{Updating theories with AI}

As demonstrated earlier with the Devil's Advocate vignette, AI is less encumbered by social considerations or cognitive dissonance when offering contrarian viewpoints.  AI can help overcome cognitive biases and organizational inertia, which often hinder effective theory updates.  LLMs could assist in recalibrating theories to better fit new data, potentially reducing the tendency to overfit or underfit models \citep{Mart16,Csas20b}.

However, the TBV would argue that excessive dependence on AI for updating theories might lead to premature strategic shifts based on misinterpreted data.  Entrepreneurs may be better at thinking outside of the box and questioning the validity of experiments \citep[\p 389]{Zell23}.  Furthermore, AI-driven updates might homogenize firms' decisions, as AI systems trained on common datasets might push companies towards similar strategies, diluting the distinctiveness that the TBV upholds as essential for competitive advantage \citep[\p 128]{Feli09}.  Additionally, the unpredictable nature of future market conditions might limit AI's effectiveness in updating theories, as past data may not reliably forecast future trends \citep[\p 1289]{Ehri22}.  Humans may be better at dealing with these novel environments \citep{Feli24}.

\subsubsection{The necessity of theories in AI-driven strategy}

An important question is whether theories in strategy are necessary for AI systems or they are merely a human construct to cope with bounded rationality.  This consideration leads to two contrasting perspectives.

On one hand, theories may be necessary even for superhuman AI.  Theories could be the most efficient way to devise experiments and incorporate feedback \citep{Agar23}, even for AI.  The scientific enterprise, arguably less boundedly rational than any individual or firm, relies heavily on theory to guide experimentation.  This suggests that AI strategists might also benefit from a theory-driven approach.  As Kurt Lewin \citeyearpar[\p 169]{Lewi52} aptly put it, ``There is nothing more practical than a good theory.''

On the other hand, theories may be unnecessary for AI.  Drawing a parallel from speech recognition, significant advancements have been achieved by prioritizing prediction over theory.  Fred Jelinek, director of IBM's speech recognition lab, famously stated, ``Every time I fire a linguist, the performance of the speech recognizer goes up'' \citep{Moor05}.  This perspective suggests that AI might make strategic decisions without relying on explicit theories, instead leveraging its vast data processing capabilities to make predictions directly \citep[see, e.g.,][]{Ludw24}.  This aligns with Shmueli's \citeyearpar{Shmu10} argument that accurate prediction can sometimes eliminate the need for theoretical explanation, and DeStefano et al.'s \citeyearpar{DeSt22} finding that providing humans with interpretable algorithms may sometimes lead to worse decisions.

It remains uncertain whether the role of theory in AI-driven strategy will align more closely with Lewin's theory-centric view or Jelinek's data-driven approach.  Regardless, the integration of AI into SDM offers a unique opportunity to explore and potentially redefine the role and necessity of theories within strategy.  This exploration may lead to new insights about the nature of strategic thinking itself, whether performed by humans or AI systems.

\subsection{Conclusion}

In the Introduction, we wondered whether SDM will follow the same path as financial trading, which 30 years ago was solely done by humans yet today is mostly conducted by algorithms \citep{SEC20}.  Will SDM be mostly handled by AIs 30 years from now?  The answer will depend on how quickly AI will continue to advance, which depends on scientific and technological breakthroughs that are hard to predict.  But even if AI development were to plateau at its current level, it is likely that SDM would still need to adapt.  Today's AI is already capable of performing useful analyses, such as those illustrated in sections~\ref{sec:how} and~\ref{sec:evidence}, and potentially some of the analyses outlined in Section~\ref{sec:implications}.  We have only begun to explore the possibilities at the intersection of AI and strategy; hence, extensive further exploration is needed.

A more general opportunity that AI opens up for strategy is that AI enables strategy ideas and frameworks to be executed by machines.  This capability is significant because, as \citet{Csas20} suggests, if an idea can be executed by a machine, it is likely that it is well understood.\footnote{This could be because humans were able to successfully program the solution to the problem, or because the machine learned the solution on its own, which then allowed humans to conduct detailed analyses or train alongside it (see \citealt{Gaes23} for an example of the latter).}  That is, moving from strategy frameworks to strategy algorithms would be accompanied by a drastic increase in the precision of ideas in strategy.  Which, following Knuth's \citeyearpar[\p 668]{Knut74} idea that ``science is knowledge which we understand so well that we can teach it to a computer,'' would imply that AI could be the defining factor in giving rise to a ``strategy science.''

The scenarios and implications resulting from the integration of AI into SDM that we have discussed throughout this paper suggest that this integration is likely the most significant change in the history of SDM.  This is because this integration has the potential to fundamentally alter the processes underlying SDM---search, representation, and aggregation---as well as its primary outcomes, the generation and evaluation of strategies.  As AI continues to advance, such changes have the potential to boost the quality, efficiency, heterogeneity, and availability of SDM, making the augmentation of human strategists by AI a definite possibility.  This transformation could spark a Cambrian explosion of new ideas and research on SDM, as AI-augmented SDM propels the field into uncharted territories.  This paper has endeavored to map out the initial contours of this evolving terrain, which may shortly become central to the strategy field's intellectual landscape.  We eagerly anticipate further exploration of this new frontier.

\clearpage
\begin{singlespacing}
\bibliography{bibliography-ai-sdm}
\end{singlespacing}

\clearpage
\appendix
\setcounter{figure}{0}
\numberwithin{figure}{section}
\setcounter{table}{0}
\numberwithin{table}{section}

\section{Vignettes}\label{app:vignettes}

\subsection{Scenario Planning}\label{app:scenario-planning}

\begin{code}{Prompt:}{}
Can you generate three different future scenarios if I choose to go ahead with establishing such a business school with one billion dollars given by a donor? Consider the competitors' potential response as well. Generate a detailed analysis where you dedicate at least 100 words to each scenario. Be as creative as possible, but be realistic as well.
\end{code}

\begin{code}{Answer:}{\itshape}
Certainly! Here are three potential future scenarios for your new business school, considering a budget of one billion dollars and possible responses from competitors:

1. Scenario: "Tech-Forward Titan"

Description: In this future, your school invests heavily in a cutting-edge digital platform, state-of-the-art campus facilities, and a blended learning model. The curriculum is shaped around future business trends, including AI, blockchain, and sustainable business practices.

Competitors' Response: Seeing the success of the blended learning model, traditional institutions scramble to upgrade their digital infrastructures. They form partnerships with tech companies to boost their online presence.

Analysis: Being a tech-forward institution would set the school apart, especially in regions where such models aren't common. The initial investment in technology could be hefty, but the potential for scaling and reaching a global student audience can provide significant returns. This scenario depends on executing the blend of online and offline experiences seamlessly and continually updating tech capabilities.

2. Scenario: "The Global Networker"

Description: Instead of creating a single massive campus, you establish multiple smaller campuses in strategic locations across the globe. The focus is on experiential learning, cultural exchange, and creating a truly global business network for students.

Competitors' Response: Traditional institutions might form more international partnerships, offering exchange programs and global internships to counter your school's widespread presence.

Analysis: The distributed model taps into the globalizing trend in business education. The key challenge would be maintaining a consistent quality of education across locations. However, students would be attracted to the global exposure, making the school a magnet for those seeking international careers. The interconnected network could also attract multinational companies seeking a diverse pool of talent.

3. Scenario: "Incubator of Innovation"

Description: Your school adopts a unique approach where students don't just learn about business; they start one. A significant portion of the curriculum is dedicated to entrepreneurship, and the billion-dollar fund is partly used to finance student startups.

Competitors' Response: Seeing the real-world success of your graduates, competing schools bolster their entrepreneurship programs. They establish more ties with venture capitalists and angel investors to facilitate student startups.

Analysis: This approach makes the school an epicenter for innovation and attracts budding entrepreneurs worldwide. It also creates a self-sustaining ecosystem where successful alumni can reinvest in the school or fund future student ventures. The challenge lies in providing sufficient support, mentorship, and resources to ensure the success of these startups. Additionally, a strong network with the business community would be crucial.

Each scenario has its unique benefits and challenges, but with a billion-dollar budget, the potential to reshape the business education landscape is immense. The key lies in effective execution, continual adaptation, and staying ahead of competitor responses.
\end{code}

\subsection{Porter's Five Forces}\label{app:five-forces}

\begin{code}{Prompt:}{}
I am the Dean of a new business school (to be established this year) which is seeking to disrupt business school education and create the business school of the future. Can you perform a Porter's Five Forces Analysis for the business school education industry? Generate a detailed analysis where you dedicate at least 100 words to each industry force.
\end{code}

\begin{code}{Answer:}{\itshape}
Certainly! Porter's Five Forces Analysis is a framework that helps to analyze the level of competition within an industry and business strategy development. Here's an analysis of the business school education industry based on this framework:

1. Threat of New Entrants (Barrier to Entry) The business education sector has seen a surge in the number of institutions due to the increasing demand for business skills worldwide. The primary barriers to entry include high startup costs (infrastructure, faculty recruitment, accreditation processes) and the reputation & brand value of established institutions. New entrants might struggle to attain the prestige of well-known business schools. However, the rise of online education platforms is lowering barriers to entry, enabling institutions to offer courses without heavy infrastructure costs. The use of technology and innovative teaching methodologies can potentially disrupt traditional models, paving the way for newer institutions.

2. Bargaining Power of Buyers (Students) With an increase in the number of business schools, students have more choices now than ever. This gives them substantial bargaining power. Schools compete for top students, often offering scholarships, innovative programs, and other incentives. Students are increasingly considering factors like employability after graduation, curriculum relevance, and return on investment when choosing schools. The rise of online courses and MOOCs also provides students with alternative, often cheaper, options for business education. Consequently, schools must constantly innovate and adapt to the changing needs and expectations of students.

3. Bargaining Power of Suppliers (Faculty and Staff) Top-quality faculty are integral to the reputation and effectiveness of a business school. With a limited pool of top-tier faculty, especially in specialized areas, they possess significant bargaining power. Faculty demand competitive salaries, research grants, and other benefits, especially if they have offers from other prestigious institutions. The supplier power is especially high for renowned professors or those who have achieved significant milestones in their field. The importance of reputation in academic circles cannot be overstated, and institutions that want to be recognized must be willing to compete for the best talent.

4. Threat of Substitute Products or Services Substitutes in this context include alternative education models like online courses, vocational training, certification programs, corporate in-house training, and self-study platforms. With the rise of digital technology, there's been a surge in online platforms like Coursera, edX, and Udemy offering business courses, sometimes in collaboration with top-tier universities. These alternatives can be more flexible, affordable, and accessible than traditional business school programs. Institutions need to offer unique value propositions or integrate these new models into their curriculum to remain relevant.

5. Rivalry among Existing Competitors The competition among business schools is fierce. Traditional top-tier institutions compete for the same pool of students, faculty, and corporate partners. Rankings by various publications exacerbate this competition, as institutions strive to outdo each other in metrics like research, student employability, and alumni earnings. The intensity of rivalry drives innovation in curriculum design, pedagogy, and extracurricular offerings. Newer entrants, on the other hand, often seek niche areas or unique value propositions, like focusing on entrepreneurship, digital transformation, or social impact, to differentiate themselves.

In conclusion, the business school landscape is undergoing significant change, driven by technological advancements and evolving student needs. For a new school seeking to disrupt the industry, understanding these forces can be invaluable in crafting a unique and compelling strategy.
\end{code}

\subsection{Devil's Advocate}\label{app:devils-advocate}

\begin{code}{Prompt:}{}
I am the Dean of a new business school (to be established this year) which is seeking to disrupt business school education and create the business school of the future. Assume that I have one billion dollars from a donor to spend on setting up this business school. Based on our analysis, we have decided to set up a tech and entrepreneurship-focused business school. We plan to invest heavily in a cutting-edge digital platform, state-of-the-art campus facilities, and a blended learning model. The curriculum is shaped around future business trends, including AI, blockchain, and sustainable business practices. Furthermore, students don't just learn about business; they start one. A significant portion of the curriculum is dedicated to entrepreneurship, and the billion-dollar fund is partly used to finance student startups. Take the role of a devil's advocate and provide arguments against this approach. Be as detailed as possible, there is no word limit.
\end{code}

\begin{code}{Answer:}{\itshape}
Certainly. Playing the devil's advocate allows us to identify potential pitfalls and challenges that might arise. Here are some arguments against the approach you've outlined for the business school:

1. Overemphasis on Tech & Entrepreneurship: While technology and entrepreneurship are undoubtedly important and will shape the future of business, there's a risk that an overemphasis on these areas may lead to a narrow focus. Traditional disciplines such as finance, strategy, operations, and organizational behavior still play foundational roles in shaping businesses and shouldn't be overlooked.

2. Volatile Nature of Technology: The tech industry is known for rapid changes, and what's groundbreaking today might be obsolete in a few years. There's a potential risk of the curriculum being too reactive to current trends and not being adaptable for the long term.

3. High Failure Rate of Startups: Most startups fail. Using a significant portion of the billion-dollar fund to finance student startups may not yield the desired ROI. This might lead to concerns about the sustainability of the business school's financial model.

4. Resource Allocation Challenges: With such a significant endowment, there could be challenges regarding efficient resource allocation. A state-of-the-art campus and cutting-edge digital platforms can be expensive to maintain and update.

5. Mismatch of Student Expectations: Some students may be looking for traditional MBA learning experiences, networking opportunities, or transitioning to different careers. The proposed model may not cater to these needs, limiting the diversity and richness of the student body.

6. Faculty Recruitment & Retention: It might be challenging to find faculty with the right blend of entrepreneurial experience, tech expertise, and academic credentials. Further, the constantly evolving tech landscape might make it tough to retain them, especially if industry offers more lucrative opportunities.

7. Quality Control: With an emphasis on entrepreneurship and possibly numerous startups being launched simultaneously, it might be difficult to maintain quality control and ensure that all projects receive the required attention and mentorship.

8. Blended Learning Model Limitations: While blended learning can be beneficial, not all students thrive in such environments. Some may prefer more traditional or more fully online approaches, and the institution might lose out on this demographic.

9. Regulatory & Accreditation Challenges: As a disruptive institution, the business school may face challenges obtaining necessary accreditation or adhering to regulatory standards set for traditional business schools.

10. Reputational Risk: Being on the cutting edge means taking risks. If some of these ventures or teaching methodologies fail, it could harm the school's reputation.

11. Ethical Concerns: Investing in student-led startups could lead to potential conflicts of interest. How will the school decide which startups to fund? Will there be preferential treatment or biases?

12. Balancing Practical and Theoretical Learning: While hands-on experience is invaluable, a deep understanding of theoretical frameworks is also crucial. Striking a balance may be challenging, leading to graduates who might be lacking in either practical or theoretical knowledge.

13. Sustainability Questions: An emphasis on sustainable business practices is commendable. However, there's a risk that other core business disciplines might be sacrificed for this. Further, defining what "sustainability" means in practice and ensuring it's adequately integrated across the curriculum could be complex.

14. Long-Term Viability: Business trends evolve. What seems like a future-focused curriculum today might not be relevant in 10 or 20 years. Constantly adapting to the ever-evolving business world might be draining on resources.

15. Overreliance on the Endowment: With a billion dollars, there might be a temptation to rely heavily on this fund. If the school doesn't establish other revenue streams or allocate resources judiciously, it might face financial challenges in the future.

In conclusion, while the intention to disrupt the business education landscape is commendable, there are several potential pitfalls and challenges to be wary of. Balancing innovation with tried-and-true methods, ensuring diverse student needs are met, and planning for long-term sustainability are crucial.
\end{code}

\section{Generation study}\label{app:generation}

For this experiment, we obtained business plans submitted to a startup accelerator program based in Europe in 2021--2022.  We obtained the full set of 309 submissions, 27 of which were accepted.  We screened out business plans that were in the local language or with any incomplete fields without written text.  This process resulted in 17 accepted business plans and 142 rejected business plans from the accelerator.

We selected a subset of 5 accepted and 5 rejected business plans from this set. We first trimmed the 17 accepted business plans by 70\% according to word count, which provided us with a list of 12 accepted business plans with more representative lengths between 629 and 2140 words.  We conducted the same trimming on the rejected business plans, which resulted in 100 business plans with word counts from 378 to 1455, 46 of which were rejected in the first screening round by accelerator staff with prior venture capital or industry experience based on the business plan alone.  We then randomly selected 5 accepted and 5 rejected business plans.

We created LLM-generated versions of these selected business plans.  First, we fixed any formatting issues in business plans using the following prompt:

\begin{code}{Prompt to fix formatting:}{}
Fix the typos in the following text and also fix the formatting so that it complies with Markdown. Do not change the content, just fix typos and Markdown formatting. When converting to Markdown, note that the first-level headers in the document are in all UPPERCASE.
\end{code}

We then used the following prompt to create the LLM-generated versions of business plans, followed by the first section of the original business plan:

\begin{code}{Prompt to complete a business plan:}{}
The following is the first section of a business plan. Write the rest of the business plan in Markdown format. The first-level headers for the rest of the business plan should be: SOLUTION, MARKET, GO-TO-MARKET STRATEGY.
\end{code}

To ensure that LLM-generated versions were of similar lengths to the original business plans, we created several alternative business plans and chose the one closest in the number of words. To create these alternative business plans, we used the following prompts:

\begin{code}{Prompt to shorten a business plan:}{}
The following is the business plan of a start-up firm. Write a slightly shorter version of it, while trying to keep all the important information. This new version should also be written in Markup format and have all the same first-level headers.
\end{code}

\begin{code}{Prompt to enlarge a business plan:}{}
Below you'll find the business plan of a start-up firm. Write a 50% longer and more detailed version of this business plan. This new version should also be written in Markup format and have all the exact same first-level headers as the original business plan, that is: PROBLEM, SOLUTION, MARKET, and GO-TO-MARKET STRATEGY.

To write this more detailed business plan, write in a more verbose style and add details about aspects that venture capitalist would like to know (e.g., details about the team, execution plan, financials, structure of venture summary, innovation & value, proposition, market opportunity & strategy, sustainable competitive advantage, product &, services, potential to invest in this idea, and viability). Feel free to add more arguments, nuance, and flair to the business plan. The new business needs to be considerably longer than the original one.
\end{code}

\begin{figure}
\caption{Rubric used by evaluators.}\label{fig:rubric}
{\centering \includegraphics[width=0.6\linewidth]{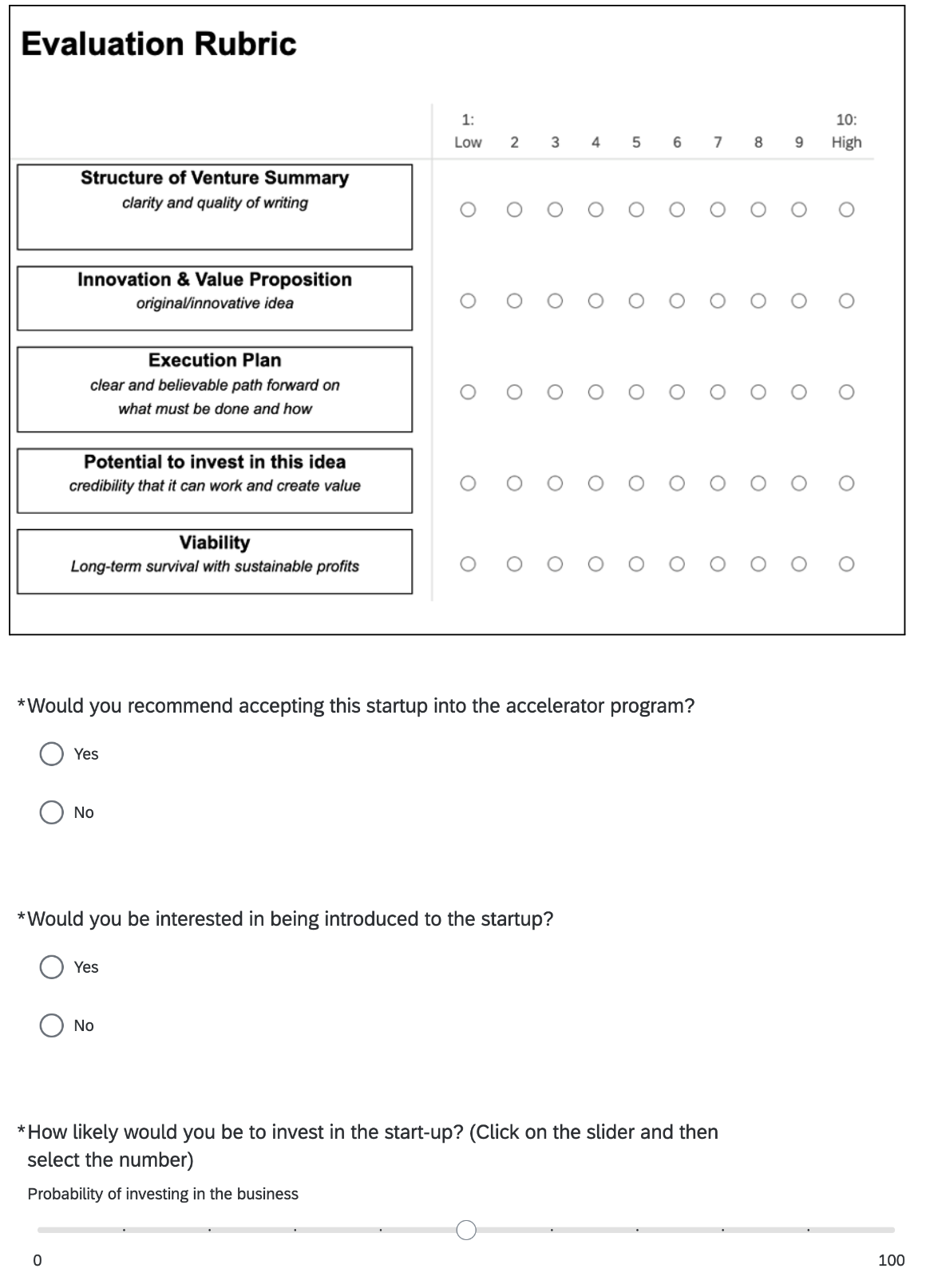}\par}
{\footnotesize \emph{Notes.}  This figure shows the scoring rubric that evaluators were asked to complete for each business plan that they assessed.}
\end{figure}

We randomly assigned evaluators to one of the versions of each business plan (the original or LLM-generated) for the set of ten plans, and additionally randomized the display order.  All the randomization was coded into a Qualtrics survey. Figure \ref{fig:rubric} shows the rubric used by evaluators.

Evaluators were recruited online through a survey panel, screened for geographic location (United States) and experience in angel or venture capital investing.

Table \ref{tbl:sumstats} shows all descriptive statistics of our sample.

Table \ref{tbl:factors} compares evaluator scores for entrepreneur- versus LLM-generated plans by factor and aggregated into an index.

\begin{table}[b]
\begin{center}
\caption{Evaluators' descriptive statistics.}\label{tbl:sumstats}
\def\sym#1{\ifmmode^{#1}\else\(^{#1}\)\fi}
\begin{tabular}{l*{5}{ccccccc}}
\hline\hline
                    &\multicolumn{1}{c}{Mean}&\multicolumn{1}{c}{SD}&\multicolumn{1}{c}{ Min.}&\multicolumn{1}{c}{Max.}&\multicolumn{1}{c}{Count}\\
                    &            &            &            &            &            \\
\hline
Years of investment experience&        4.60&        4.66&        0.00&       35.00&         250\\
Number of investments&        3.70&        4.26&        0.00&       30.00&         250\\
Experience - angel investing&        0.37&        0.48&        0.00&        1.00&         250\\
Experience - venture capital&        0.48&        0.50&        0.00&        1.00&         250\\
Experience - private equity&        0.62&        0.49&        0.00&        1.00&         250\\
Years of work experience&       15.76&        9.04&        1.00&       30.00&         250\\
Years of managerial experience&        6.78&        6.78&        0.00&       30.00&         250\\
Degree higher than college&        0.39&        0.49&        0.00&        1.00&         250\\
Female              &        0.34&        0.47&        0.00&        1.00&         250\\
\hline\hline
\end{tabular}
\\
\end{center}
{\footnotesize \emph{Notes.}  \textit{Experience} in angel investing, venture capital, and private equity were captured as binary variables. \textit{Degree higher than college} and \textit{Female} are also coded as binary variables.}
\end{table}

\begin{table}
\begin{center}
\caption{Comparison of entrepreneur- vs.\ LLM-generated business plans by factor.}\label{tbl:factors}
\def\sym#1{\ifmmode^{#1}\else\(^{#1}\)\fi}
\begin{tabular}{l*{6}{c}}
\hline\hline
                     &\multicolumn{1}{c}{(1)}&\multicolumn{1}{c}{(2)}&\multicolumn{1}{c}{(3)}&\multicolumn{1}{c}{(4)}&\multicolumn{1}{c}{(5)}&\multicolumn{1}{c}{(6)}\\
                     &\multicolumn{1}{c}{Writing}&\multicolumn{1}{c}{Innovation}&\multicolumn{1}{c}{Execution}&\multicolumn{1}{c}{Investment}&\multicolumn{1}{c}{Viability}&\multicolumn{1}{c}{Evaluation}\\
                    &\multicolumn{1}{c}{Quality}&\multicolumn{1}{c}{}&\multicolumn{1}{c}{}&\multicolumn{1}{c}{Potential}&\multicolumn{1}{c}{}&\multicolumn{1}{c}{Index}\\
\hline
LLM                 &        0.23\sym{***}&        0.02         &        0.23\sym{***}&        0.09\sym{***}&        0.07\sym{**} &        0.14\sym{***}\\
                    &      (0.03)         &      (0.03)         &      (0.03)         &      (0.03)         &      (0.03)         &      (0.03)         \\
Plan FE             &         Yes         &         Yes         &         Yes         &         Yes         &         Yes         &         Yes         \\
Investor FE         &         Yes         &         Yes         &         Yes         &         Yes         &         Yes         &         Yes         \\
\hline
Observations        &        2500         &        2500         &        2500         &        2500         &        2500         &        2500         \\
\hline\hline

\end{tabular}

\end{center}
{\footnotesize \emph{Notes.}  Each column from (1) to (5) represents one of five factors of the business plan evaluated on a scale of 1 (low) to 10 (high): (1) writing quality; (2) innovation and value proposition; (3) execution plan; (4) potential to invest in the idea; (5) viability, all standardized.  Column (6) indicates an index variable coded as a standardized sum of all five factors.  Standard errors are reported in parentheses.  * $p<0.10$, ** $p<0.05$, *** $p<0.01$.}
\end{table}

\FloatBarrier
\section{Evaluation study}\label{app:evaluation}

The following sequence of three prompts were used to compute the AI scores reported in Section~\ref{sec:evaluation}.

\begin{code}{Prompt 1:}{}
You are an experienced venture capitalist. You have also written a textbook on how to make successful investments in startup companies and have taught a course based on it in the MBA program. Your textbook builds on practical experience as well as academic research developed in the fields of strategy and entrepreneurship.

Soon you'll have to participate in a business plan competition. For each company, you'll read a document summarizing its business plan and then you will have to provide scores (from 0% to 100%) for each of the following 10 dimensions:

1. Team (e.g., quality of team and fit with the idea, advisory board, strategy partners)

2. Execution Plan (e.g., clear and believable path forward, they know what must be done and how, timeline and execution plan)

3. Financials (e.g., business model, financial forecasts & valuation, funding required & equity offered)

4. Structure of Venture Summary (e.g., clarity and quality of writing, structure and sequence of ideas, the document reads well)

5. Innovation & Value Proposition (e.g., original/innovative idea, solving a problem for customers, must-have vs. nice-to-have, feasible, marketable, profitable)

6. Market Opportunity & Strategy (e.g., potentially large/fast growing market, niche identification, competitor and customer analysis, effective strategy)

7. Sustainable Competitive Advantage (e.g., sustainability, differentiation from competitors, IP strategy)

8. Product & Services (e.g., clear product/service/platform description, product/service is superior to competitors, addresses actual customer needs)

9. Potential to Invest in this Idea (e.g., credibility and persuasiveness, it can work and will create value, understanding of entrepreneurship, demonstrates industry knowledge)

10. Viability (e.g., long-term survival, sustainable profits)

From the book you wrote, these are the types of concepts that have in mind when scoring startups along each of the 10 dimensions:

1. Team: The textbook emphasizes on assessing the competence and experience of the team along with their synergy and fit with the venture. Also, the presence of an effective advisory board, and strategic partners denotes strong networks which is crucial for start-ups.

2. Execution Plan: The textbook would underpin the need for a clear, believable, and realistically achievable execution plan underlined by discernible action steps and timeline. Understanding how they plan to deal with potential obstacles and challenges is also important.

3. Financials: The textbook advises to assess the business model, margin analysis, profit forecast, projected cash flow, the need for funding, value assigned, and equity offered among others. The financial strategy should reflect understanding of the market need and scalability of the product/service.

4. Structure of Venture Summary: The textbook stresses on the need for clarity, quality of writing, logical flow of ideas, and comprehensiveness of the summary presentation.

5. Innovation & Value Proposition: The textbook underscores the importance of originality, novelty, ability to solve a real problem, the feasibility, marketability, and profitability of the idea as well as its positioning as a must-have or nice-to-have in the market.

6. Market Opportunity & Strategy: The textbook advises to evaluate the market size and growth prospects, identification of market niche, depth of competitor and customer analysis, and effectiveness of GTM (Go-To-Market) strategy.

7. Sustainable Competitive Advantage: The textbook advises to examine differentiation from competitors, barriers to entry, Intellectual Property Rights strategy for ensuring exclusivity and sustainability.

8. Product & Services: The textbook emphasizes on the clarity of the product/service/platform, superiority to competitors, and alignment with actual customer needs/ wants.

9. Potential to Invest in this Idea: The textbook stresses on credibility and persuasiveness of the pitch, feasibility of the idea, understanding of entrepreneurship, demonstration of industry knowledge, and return on investment potential.

10. Viability: The textbook advises to assess the capacity for long-term survival, the ability to generate sustainable profits, adaptability in market, and overall business resilience.

Below you'll see the first startup you have to evaluate. Read it carefully and provide the main pros and cons for each dimension. Format your response as follows:

1. Team
 - Pros: ...
 - Cons: ...
2. Execution Plan
 - Pros: ...
 - Cons: ...
3. Financials
 - Pros: ...
 - Cons: ...
4. Structure of Venture Summary
 - Pros: ...
 - Cons: ...
5. Innovation & Value Proposition
 - Pros: ...
 - Cons: ...
6. Market Opportunity & Strategy
 - Pros: ...
 - Cons: ...
7. Sustainable Competitive Advantage
 - Pros: ...
 - Cons: ...
8. Product & Services
 - Pros: ...
 - Cons: ...
9. Potential to Invest in this Idea
 - Pros: ...
 - Cons: ...
10. Viability
 - Pros: ...
 - Cons: ...

The business plan to be evaluated follows.

[BUSINESS PLAN]
\end{code}

\begin{code}{Prompt 2:}{}
Given your analyses, what are the two strongest and two weakest dimensions for this startup? Why?
\end{code}

\begin{code}{Prompt 3:}{}
Given all your analyses above, now provide a score (in the 0-100% range) for each of the dimensions. For reference, a 0% is very low, a 50% is average, and 100% very high.

Format your answer as a table with the following format:

| Dimension                            | Percentage |
|--------------------------------------+------------|
| 1. Team                              |            |
| 2. Execution Plan                    |            |
| 3. Financials                        |            |
| 4. Structure of Venture Summary      |            |
| 5. Innovation & Value Proposition    |            |
| 6. Market Opportunity & Strategy     |            |
| 7. Sustainable Competitive Advantage |            |
| 8. Product & Services                |            |
| 9. Potential to Invest in this Idea  |            |
| 10. Viability                        |            |

Do not include anything else in your answer apart from the table.

\end{code}

\begin{figure}[h]
\begin{center}
\caption{Correlations between investor and LLM evaluations by training window.}\label{fig:scatterplot-pre-post-training}
\includegraphics[width=0.6\linewidth]{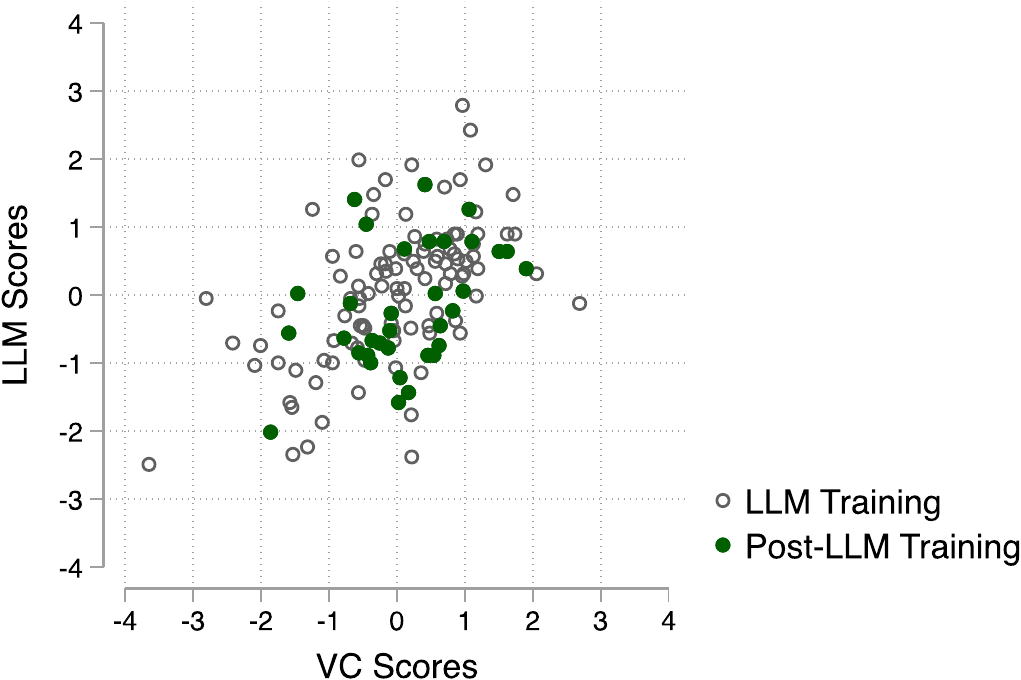}
\end{center}
{\footnotesize \emph{Notes.}  This figure shows a scatterplot of all standardized investor and LLM evaluations of 138 business plans submitted to a startup competition, highlighting plans that were submitted after the LLM training cutoff.}
\end{figure}

Figure~\ref{fig:scatterplot-pre-post-training} shows that the correlation between the human and LLM evaluations remains robust across business plans submitted before and after the LLM's training window.

\bigskip

\end{document}